\documentclass[aoas,preprint]{imsart}
\setattribute{journal}{name}{}

\RequirePackage[OT1]{fontenc}
\RequirePackage[authoryear]{natbib}
\startlocaldefs

\usepackage[latin9]{inputenc}
\usepackage{float}

\usepackage{amsthm}
\usepackage{amssymb,mathrsfs}
\usepackage{color, graphicx}
\usepackage{amsbsy}
\usepackage{epsfig}
\usepackage{enumerate}
\usepackage{bm}

\makeatletter

\newcommand{\code}[1]{\texttt{#1}}

\newcommand{\ignore}[1]{}  

\renewcommand{\baselinestretch}{1.5}
\font\twelvesmc=cmcsc10 scaled\magstep1 
\newcommand{\smc}{\twelvesmc}
\newtheorem{lm}{ \underline{\smc Lemma}}

\newtheorem{pp}[lm]{{\smc Proposition}}

\newtheorem{assumption}[lm]{{\smc Assumption}}

\newcommand{\bd}{\begin{document}}
\def\qed{\hfill$\diamondsuit$}
\newcommand{\ed}{\end{document}}
\def\bse{\begin{eqnarray*}}
\def\ese{\end{eqnarray*}}
\def\be{\begin{eqnarray}}
\def\ee{\end{eqnarray}}
\newcommand{\ben}{\begin{eqnarray}}\newcommand{\een}{\end{eqnarray}}
\newcommand{\bn}{\begin{enumerate}}
\newcommand{\en}{\end{enumerate}}
\newcommand{\im}{\item}
\newcommand{\bc}{\begin{cases}}
\newcommand{\ec}{\end{cases}}
\newcommand{\bt}{\begin{tabular}}
\newcommand{\et}{\end{tabular}}
\newcommand{\bct}{\begin{center}}
\newcommand{\ect}{\end{center}}
\def\wt{\widetilde}
\def\diag{\hbox{diag}}
\def\wh{\widehat}
\def\AIC{\hbox{AIC}}
\def\BIC{\hbox{BIC}}
\newcommand{\tr}{{\rm tr}}
\newcommand{\cid}{\buildrel d \over \longrightarrow}
\newcommand{\cas}{\buildrel a.s \over \longrightarrow}
\newcommand{\cw}{\buildrel w \over \longrightarrow}
\newcommand{\cip}{\buildrel p \over \longrightarrow}
\newcommand{\cil}{\buildrel L^2 \over \longrightarrow}
\newcommand{\bl}{{\pmb l}}
\newcommand{\bk}{{\pmb k}}
\newcommand{\bda}{{\pmb a}}
\newcommand{\bdb}{{\pmb b}}
\newcommand{\bdc}{{\pmb c}}
\newcommand{\bdd}{{\pmb d}}
\newcommand{\bdf}{{\bf f}}
\newcommand{\bdg}{{\bf g}}
\newcommand{\bds}{{\pmb s}}
\newcommand{\bdt}{{\pmb t}}
\newcommand{\bdu}{{\bf u}}
\newcommand{\bdv}{{\bf v}}
\newcommand{\bw}{{\pmb w}}
\newcommand{\bdx}{{\pmb x}}
\newcommand{\bdy}{{\pmb y}}

\newcommand{\bone}{{\pmb 1}}
\newcommand{\bdepsilon}{{\pmb\epsilon}}
\newcommand{\bdeta}{{\pmb\eta}}
\newcommand{\bdmu}{{\pmb \mu}}
\newcommand{\bdpsi}{{\pmb \psi}}
\newcommand{\bdalpha}{{\pmb \alpha}}
\newcommand{\bdbeta}{{\pmb \beta}}
\newcommand{\bdgamma}{{\pmb \gamma}}
\newcommand{\bdlambda}{{\pmb \lambda}}
\newcommand{\bdtheta}{{\pmb \theta}}
\newcommand{\bddelta}{{\pmb \delta}}
\newcommand{\bdsigma}{{\pmb \sigma}}
\newcommand{\bdomega}{{\pmb \omega}}
\newcommand{\bdpi}{{\pmb \pi}}
\newcommand{\bdvartheta}{{\pmb \vartheta}}

\newcommand{\CA}{{\cal A}}
\newcommand{\CB}{{\cal B}}
\newcommand{\CC}{{\cal C}}
\newcommand{\CD}{{\cal D}}
\newcommand{\CE}{{\cal E}}
\newcommand{\CF}{{\cal F}}
\newcommand{\CG}{{\cal G}}
\newcommand{\CH}{{\cal H}}
\newcommand{\CK}{{\cal K}}
\newcommand{\CL}{{\cal L}}
\newcommand{\CM}{{\cal M}}
\newcommand{\CN}{{\cal N}}
\newcommand{\CO}{{\cal O}}
\newcommand{\CP}{{\cal P}}
\newcommand{\CR}{{\cal R}}
\newcommand{\CS}{{\cal S}}
\newcommand{\CT}{{\cal T}}
\newcommand{\CV}{{\cal V}}
\newcommand{\CW}{{\cal W}}
\newcommand{\CX}{{\cal X}}
\newcommand{\CY}{{\cal Y}}
\newcommand{\CZ}{{\cal Z}}
\newcommand{\BR}{{\bold R}}
\newcommand{\BA}{{\pmb A}}
\newcommand{\BB}{{\pmb B}}
\newcommand{\BI}{{\pmb I}}
\newcommand{\BL}{{\pmb L}}
\newcommand{\BS}{{\pmb S}}
\newcommand{\BT}{{\pmb T}}
\newcommand{\BW}{{\pmb W}}
\newcommand{\BX}{{\pmb X}}
\newcommand{\BV}{{\pmb V}}
\newcommand{\BZ}{{\bold Z}}
\newcommand{\BY}{{\pmb Y}}
\newcommand{\FC}{\field{C}}
\newcommand{\la}{{\langle}}
\newcommand{\ra}{{\rangle}}
\newcommand{\E}{{\rm E}}

\newcommand{\BCI}{{\pmb {\cal I}}}
\newcommand{\BCT}{\boldsymbol{\mathcal{T}}}

\newcommand{\MSL}{\mathscr{L}}
\newcommand{\MSR}{\mathscr{R}}

\newcommand{\el}{L^2[0,1]}
\def\II{I\negthinspace I}
\def\III{I\negthinspace I\negthinspace I}
\renewcommand{\baselinestretch}{1.2}
\newcommand{\bepsilon}{{\pmb varepsilon}}
\newcommand{\ep}{\vskip-.3cm\noindent\begin{flushright}
${{\hfill\llap{$\sqcup\!\!\!\!\sqcap$}}}$
\end{flushright}}

\def\wt{\widetilde}
\def\diag{\hbox{diag}}
\def\wh{\widehat}
\def\AIC{\hbox{AIC}}
\def\BIC{\hbox{BIC}}
\newcommand{\Appendix}{
\def\thesection{Appendix~\Alph{section}}
\def\thesubsection{A.\arabic{subsection}}
}
\def\diag{\hbox{diag}}
\def\log{\hbox{log}}
\def\bias{\hbox{bias}}
\def\sd{\hbox{sd}}
\def\Siuu{\boldSigma_{i,uu}}
\def\ANNALS{{\it Annals of Statistics}}
\def\BIOK{{\it Biometrika}}
\def\whT{\widehat{\Theta}}
\def\STATMED{{\it Statistics in Medicine}}
\def\STATSCI{{\it Statistical Science}}
\def\JSPI{{\it Journal of Statistical Planning \& Inference}}
\def\JRSSB{{\it Journal of the Royal Statistical Society, Series B}}
\def\BMCS{{\it Biometrics}}
\def\COMMS{{\it Communications in Statistics, Theory \& Methods}}
\def\JQT{{\it Journal of Quality Technology}}
\def\STIM{{\it Statistics in Medicine}}
\def\TECH{{\it Technometrics}}
\def\AJE{{\it American Journal of Epidemiology}}
\def\JASA{{\it Journal of the American Statistical Association}}
\def\CDA{{\it Computational Statistics \& Data Analysis}}
\def\JCGS{{\it Journal of Computational and Graphical Statistics}}
\def\JCB{{\it Journal of Computational Biology}}
\def\BIOINF{{\it Bioinformatics}}
\def\JAMA{{\it Journal of the American Medical Association}}
\def\JNUTR{{\it Journal of Nutrition}}
\def\JCGS{{\it Journal of Computational and Graphical Statistics}}
\def\LETTERS{{\it Letters in Probability and Statistics}}
\def\JABES{{\it Journal of Agricultural, Biological and
                      Environmental Statistics}}
\def\JASA{{\it Journal of the American Statistical Association}}
\def\ANNALS{{\it Annals of Statistics}}
\def\JSPI{{\it Journal of Statistical Planning \& Inference}}
\def\TECH{{\it Technometrics}}
\def\BIOK{{\it Bio\-me\-tri\-ka}}
\def\JRSSB{{\it Journal of the Royal Statistical Society, Series B}}
\def\BMCS{{\it Biometrics}}
\def\COMMS{{\it Communications in Statistics, Series A}}
\def\JQT{{\it Journal of Quality Technology}}
\def\SCAN{{\it Scandinavian Journal of Statistics}}
\def\AJE{{\it American Journal of Epidemiology}}
\def\STIM{{\it Statistics in Medicine}}
\def\ANNALS{{\it Annals of Statistics}}
\def\whT{\widehat{\Theta}}
\def\STATMED{{\it Statistics in Medicine}}
\def\STATSCI{{\it Statistical Science}}
\def\JSPI{{\it Journal of Statistical Planning \& Inference}}
\def\JRSSB{{\it Journal of the Royal Statistical Society, Series B}}
\def\BMCS{{\it Biometrics}}
\def\COMMS{{\it Communications in Statistics, Theory \& Methods}}
\def\JQT{{\it Journal of Quality Technology}}
\def\STIM{{\it Statistics in Medicine}}
\def\TECH{{\it Technometrics}}
\def\AJE{{\it American Journal of Epidemiology}}
\def\JASA{{\it Journal of the American Statistical Association}}
\def\CDA{{\it Computational Statistics \& Data Analysis}}
\def\dfrac#1#2{{\displaystyle{#1\over#2}}}
\def\VS{{\vskip 3mm\noindent}}
\def\boxit#1{\vbox{\hrule\hbox{\vrule\kern6pt
          \vbox{\kern6pt#1\kern6pt}\kern6pt\vrule}\hrule}}
\def\refhg{\hangindent=20pt\hangafter=1}
\def\refmark{\par\vskip 2mm\noindent\refhg}
\def\naive{\hbox{naive}}
\def\itemitem{\par\indent \hangindent2\parindent \textindent}
\def\var{\hbox{var}}
\def\cov{\hbox{cov}}
\def\corr{\hbox{corr}}
\def\trace{\hbox{trace}}
\def\refhg{\hangindent=20pt\hangafter=1}
\def\refmark{\par\vskip 2mm\noindent\refhg}
\def\Normal{\hbox{Normal}}
\def\povr{\buildrel p\over\longrightarrow}
\def\ccdot{{\bullet}}
\def\bse{\begin{eqnarray*}}
\def\ese{\end{eqnarray*}}
\def\be{\begin{eqnarray}}
\def\ee{\end{eqnarray}}
\def\bq{\begin{equation}}
\def\eq{\end{equation}}
\def\bse{\begin{eqnarray*}}
\def\ese{\end{eqnarray*}}
\def\pr{\hbox{pr}}
\def\wh{\widehat}
\def\trans{^{\rm T}}
\def\myalpha{{\cal A}}
\def\th{^{th}}
\def\wi{{\hbox{\scriptsize WI}}}

\def\boxit#1{\vbox{\hrule\hbox{\vrule\kern6pt
          \vbox{\kern6pt#1\kern6pt}\kern6pt\vrule}\hrule}}
\def\licomment#1{\vskip 2mm\boxit{\vskip 2mm{\color{black}\bf#1} {\color{blue}\bf -- Yehua\vskip 2mm}}\vskip 2mm}
\def\pancomment#1{\vskip 2mm\boxit{\vskip 2mm{\color{black}\bf#1} {\color{blue}\bf -- Lanfeng\vskip 2mm}}\vskip 2mm}

\makeatother

\endlocaldefs

\begin{document}

\begin{frontmatter}

\title{Latent Gaussian Mixture Models for Nationwide Kidney Transplant Center Evaluation}
\runtitle{Latent Gaussian Mixture Models}

\begin{aug}
\author{\fnms{Lanfeng} \snm{Pan}\thanksref{d1}\ead[label=e1]{pan@iastate.edu}},
\author{\fnms{Yehua} \snm{Li}\thanksref{d1,m3}\corref{}\ead[label=e2]{yehuali@iastate.edu}},
\author{\fnms{Kevin} \snm{He}\thanksref{d2}\ead[label=e3]{}},
\author{\fnms{Yanming} \snm{Li}\thanksref{d2}\ead[label=e4]{}}
\and
\author{\fnms{Yi} \snm{Li}\thanksref{d2}\ead[label=e5]{}}

\thankstext{m3}{Correspondence should be addressed to Yehua Li (yehuali@iastate.edu)}

\affiliation{\thanksmark{d1} Department of Statistics \& Statistical Laboratory, Iowa State University }
\affiliation{\thanksmark{d2} School of Public Health \& Kidney Epidemiology and Cost Center, University of Michigan, Ann Arbor}
\end{aug}

\begin{abstract}
Five year post-transplant survival rate is an important indicator on quality of care delivered  by kidney transplant centers in the United States. To provide a fair assessment of each transplant center,  an effect that represents the center-specific care quality, along with patient level risk factors, is often included in the risk adjustment model. In the past,  the center effects have been modeled as either fixed effects or Gaussian random effects, with various pros and cons. Our numerical analyses reveal that the distributional assumptions do impact the prediction of center effects especially when the effect is extreme. To bridge the gap between these two approaches, we propose to model the transplant center effect as a latent random variable with a finite Gaussian mixture distribution. Such latent Gaussian mixture models provide a convenient framework to study the heterogeneity among the transplant centers. To overcome the weak identifiability issues, we propose to estimate the latent Gaussian mixture model  using a penalized likelihood approach, and develop sequential locally restricted likelihood ratio tests to determine the number of components in the Gaussian mixture distribution. The fitted mixture model provides a convenient means of controlling the false discovery rate when screening for underperforming or outperforming transplant centers. The performance of the methods is verified by simulations and by the analysis of the motivating data example. 
\end{abstract}


\end{frontmatter}

\section{Introduction}\label{sec:introduction}

This paper is motivated by the analysis of the national kidney transplant data, supported in part by the Health Resources and Services Administration. Renal failure is one of the most common and severe diseases in
the nation. In 2013,  a total of 117,162 new cases  were reported (www.USRDS.org).  Kidney transplantation, as a primary therapy for end stage renal disease, 
typically involves transplant surgeons and physicians, coordinators, social workers, financial counselors,  nutritionists, psychologists, referring physicians, and the patients.  
The quality of care delivered by the transplant centers is often assessed by patient survival, for example, the 5 year survival rate post transplant.

To provide a fair assessment of each transplant center,   patient level risk factors  as well as  an effect that represents the care quality of the transplant center are often included in the risk adjustment model.  The Organ Procurement and Transplantation Network (OPTN), as a critical system in helping organ transplant institutions match waiting candidates with donated organs,
 contains all national data on the candidate waiting list, organ donation and matching, and transplantation. Kidney transplant database is a large component of OPTN, which includes the patient level risk factors such as demographical information, quality of the donor kidney, matching between the patient and the donor,  as well as  the transplant centers which operated the transplant surgeries.  It is of substantial interest to estimate transplant center effects based on this national database, as they 
  provide a data-driven basis for evaluation of national transplant centers and identification of underperforming or outperforming centers. The results
may have health-policy making implications and facilitate patients' choice of transplant centers.

Many statisticians and health policy researchers \citep{krumholz2006a, krumholzhm2006b,  li2009national}
advocate modeling the center effects as random effects that follow a Gaussian distribution. 
This approach ignores the heterogeneity among the transplant centers: there is a shrinkage effect in the prediction of the center level random effects and the assumption of a common Gaussian distribution makes the predicted random effects similar in value.  \cite{he2013a} argued that borrowing information from other transplant centers is not fair when the goal of the study is to evaluate and rank these centers. Instead, they suggested to model the transplant center effects as fixed effects. However, in such a fixed effects model, the number of parameters is large, making statistical inference numerically unstable, especially when the center size varies substantially. Estimating the effects of small centers with fewer patients presents even greater challenges. Indeed,  in our national transplant study the number of patients treated by individual centers varies from 3 to 5830, with a median center size of  603.  A comprehensive critic of these two approaches can be found in a report prepared by  the Committee of Presidents of Statistical Societies  (COPSS) through a contract with Centers for Medicare and Medicaid Services \citep{Ash2012b}.

To bridge the gap between these two approaches, we propose to model the transplant center effects using a finite Gaussian mixture model. Our model has two advantages compared to the existing models.  First, the model allows the presence of heterogeneities (e.g. the existence of clusters or subpopulations) among the transplant centers,  making it a natural framework to identify under- or out-performing centers. Second, the mixture model can be considered as a compromise between the random effects model and the fixed effects model: it reduces to the random effects model when there is only one component in the mixture distribution and it becomes the fixed effects model if each transplant center is a cluster. Within 
the framework of generalized linear mixed effects models (GLMM), we will develop data-driven methods to determine the number components in the mixture random model.

Indeed, the vast majority of the GLMM literature assumes the distribution of the random effect is Gaussian, focuses on estimating the fixed effects and treats the random effects as nuisance \citep{Breslow1993, Lin1996}. Even though GLMM is in general robust against deviation from the Gaussian random effect assumption \citep{McCulloch2011}, many authors have documented various drawbacks when the Gaussian assumption is violated, including loss of estimation efficiency \citep{Chen2002b}, reduced power for statistical tests \citep{litiere2007}, etc.  Even though the predicted random effects are relatively robust in terms of mean squared error, the shape of the distribution for the predicted random effect is highly sensitive and mostly reflects the shape of the assumed random effect distribution \citep{McCulloch2011}.
Many authors have tried to relax the Gaussian assumption and model the random effect in GLMM with more flexible distributions, such as the semi-nonparmatric distribution \citep{Chen2002b} and Gaussian mixture distribution \citep{Caffo2007}. \cite{Caffo2007} proposed a similar model as ours. However, they limited their investigation to binary probit GLMM and focused on numerical performance rather than theoretical justification. We propose a test to check if it is necessary to model random effects as normal mixture as well as how many number of components are sufficient. In addition, another major difference is our goal is random effect itself instead of relaxing the assumption on it. By modeling the random effect as normal mixture we can do an evaluation on it in a FDR way.



Finite Gaussian mixture models \citep{mclachlan2004finite} are  intuitively appealing  for modeling non-homogeneity in a population and detecting subgroup structures. There has been a recent surge in application of Gaussian mixture models, including clustering analysis \citep{Huang2014}, false discovery rate control \citep{Efron2004,Liang2008a}, genetic imprinting \citep{Li2014}. In contrast to its usefulness, however, estimation and statistical inference for Gaussian mixture models have been much difficult, because many regularity conditions in parametric inference are violated in Gaussian mixture models \citep{Hathaway1985a, Chen1995,Chen2009}.  There has been much recent work in hypothesis testing on the order of finite Gaussian mixture models \citep{Chen2012,Kasahara2014}. However, none has studied  GLMM with the  random effects modeled with  Gaussian mixtures.

The rest of the paper is organized as follows.  We introduce the model in Section \ref{sec:model} and propose an EM-based estimation procedure in Section \ref{estimation}, where the consistency of the procedure is also established. To decide the number of mixture components,
we propose sequential locally restricted likelihood ratio tests in Section \ref{sec:NUMBER-OF-COMPONENTS}. In Section \ref{sec:fdr}, we propose a false discovery rate control procedure to evaluate the care qualities of the transplant centers. We conduct simulations in Section \ref{sec:simulations} and report the analysis of the OPTN kidney transplant data in Section \ref{sec:data}.
Finally, we end the paper with concluding remarks in Section \ref{sec:summary}. A simulation procedure to evaluate the null distribution for the test statistic in Section \ref{sec:high_order_test} is provided in the appendix, and all technical proofs and additional regularity conditions are deferred to the supplementary material.

\section{Model and Assumptions}\label{sec:model}

Suppose that there are  $n$ independent transplant centers, each  treating   $N_i$ patients,
which brings the total sample size to be $N=\sum_{i=1}^n N_{i}$. 
Let $Y_{ik}$ be the outcome variable of the $k$th patient treated at the $i$th transplant center and let $\BX_{ik}\in \mathbb{R}^p$ be the patient level covariate, $k=1,2\ldots N_{i}$, $i=1,2,\ldots n$. 
Denote by  $\BY_i=(Y_{i1},\ldots, Y_{iN_i})\trans$ and $\BX_i=(\BX_{i1},\ldots, \BX_{i N_i})\trans$,
 and let $\gamma_i$ be the random effect that represents the care qualify of the $i$th center, and denote $\boldsymbol{\gamma}=(\gamma_{1},\ldots,\gamma_{n})^{T}$.
Suppose that the conditional density of $Y_{ik}$, given $\BX_{ik}$ and $\gamma_i$,  belongs to the canonical exponential family:
\ben\label{eq:glmm}
	f(Y_{ik}|\BX_{ik},\gamma_{i}; \bdbeta, \varphi)
	= \exp\bigg\{ \frac{Y_{ik} \xi_{ik} +b(\xi_{ik}) }{ a(\varphi)} +d(Y_{ik}, \varphi)\bigg\},
\een
where $a(\cdot)$, $b(\cdot)$ and $d(\cdot)$ are known functions, $\xi_{ik}=\BX_{ik}^{T}\boldsymbol{\beta}+\gamma_{i}$ is the canonical parameter with $\E(Y_{ik} | \BX_{ik},\gamma_{i}) =b'(\xi_{ik})$, and $\varphi$ is a nuisance parameter. 
We also assume that $Y_{ik} $ and $Y_{ik'}$ are independent given $\gamma_i$, for any $k\neq k'$. In our transplant center evaluation application, we consider binary response variable: $Y_{ik}=1$ if the patient deceased within 5 years after transplant; $-1$ otherwise.  In the dataset, there were essentially no censoring within the first 5 years as the transplant patients' survival information had been closely monitored and tracked. This gives the justification of treating 5 year survival as a binary outcome data. With that, model (\ref{eq:glmm}) becomes $f({Y}_{ik}|\BX_{ik},\gamma_{i};\boldsymbol{\beta})=\{1+\exp(-\xi_{ik}Y_{ik})\}^{-1}$. 


Assume that the transplant centers belong to $C$ subpopulations and the $c$th subpopulation can be described by a Gaussian distribution with mean $\mu_c$ and variance $\sigma_c^2$, $c=1,\ldots,C$. 
Marginally, the density of $\gamma_i$ is $g(\gamma |\bdtheta_{\gamma})=\sum_{c=1}^{C}\pi_{c} f_c(\gamma| \mu_c, \sigma_c)$, 
where $f_c(\gamma| \mu_c, \sigma_c)=\sigma_c^{-1} \phi\{(\gamma- \mu_{c})/\sigma_{c}\}$, $\phi (\cdot)$ is the standard Gaussian density, $\pi_c\in [0,1]$ is the weight for subpopulation $c$, $\sum_{c=1}^C \pi_c=1$, and $\bdtheta_{\gamma}=(\mu_{1},\ldots,\mu_{C},\sigma_{1},\ldots,\sigma_{C},\pi_{1},\ldots,\pi_{C})^{T}$ is the collection of parameters in $g(\gamma)$.

Denote $\BY=(\BY_1\trans,\ldots, \BY_n\trans)\trans$, $\BX=(\BX_1\trans,\ldots, \BX_n\trans)\trans$, and $\bdtheta=(\bdtheta_y\trans,\bdtheta_{\gamma}\trans)\trans$ where $\bdtheta_y=(\boldsymbol{\beta}\trans,\varphi)\trans$. To facilitate an EM algorithm, define $\boldsymbol{L}_i=({L}_{i1},\ldots,{L}_{iC})^{T} \sim$ Multinomial$(\pi_1,\ldots, \pi_C)$  as a latent random vector of subpopulation memberships, 
where $L_{ic}=1$ if $\gamma_{i}$ belongs to component $c$ and $L_{ic}=0$ otherwise. 
 Then the likelihood function for  the complete data,  comprising of both observed and latent variables, is 
\ben
	l_{comp}(\bdtheta; \BY, \BX, \bdgamma, \BL)=\sum_{i=1}^n \ell_{i, comp}(\bdtheta; \BY_i , \BX_i, \gamma_i, \BL_i), \nonumber
\een
where $\ell_{i, comp}(\bdtheta; \BY_i , \BX_i, \gamma_i, \BL_i)=\log f(\BY_{i}|\BX_{i},\gamma_{i};\bdtheta_y)+\sum_{c=1}^C L_{ic} [ \log\pi_c -\frac{1}{2} \log(\sigma_c^2)+\log \phi \{(\gamma_i-\mu_{c})/\sigma_{c}\} ] $ and $f(\BY_{i}|\BX_{i},\gamma_{i};\bdtheta_y) =\prod_{k=1}^{N_i} f(Y_{ik} | \BX_{ik}, \gamma_i; \bdtheta_y)$.

\section{Estimation Procedure}\label{estimation}

Though conceptually appealing, Gaussian mixture models possess some undesirable properties:  slower convergence rate if the number of components is unknown \citep{Chen1995}; unbounded likelihood if any of the component variance parameters $\sigma_c^2$ goes to 0 \citep{Hathaway1985a};  and infinite Fisher information on some boundary points of the parameter space \citep{Chen2009}. The solution to these problems in the literature is to either restrict the value of the parameters away from the boundaries \citep{Hathaway1985a} or include a penalty function to prevent any $\sigma_c$ from converging to 0 \citep{Chen2008a,Chen2009}. 

We propose to adopt the latter  by maximizing a penalized complete data likelihood
\ben
	l_{comp,p}(\bdtheta; \BY, \BX, \bdgamma, \BL)=l_{comp}(\bdtheta; \BY, \BX, \bdgamma, \BL)+\sum_{c=1}^Cp_n(\sigma^{2}_{c}), \label{eq:pl}
\een
while treating $\bdgamma$ and $\BL$ as missing data. \citet{Chen2008a} provided asymptotic conditions on $p_n(\sigma^2)$ that ensures the consistency of the estimator. In all of our numerical studies, we use the following penalty proposed by \cite{Chen2009}
\ben\label{eq:penalty_function}
	p_n(\sigma^{2};\wh\sigma^2_{pilot})=-a_{n}\{\wh\sigma^2_{pilot}/\sigma^{2}+\log(\sigma^{2}/\wh\sigma^2_{pilot})-1\},
\een
where $\wh\sigma^2_{pilot}$ is a pilot estimate for the variance of $\gamma$. One possible choice of $\wh\sigma^2_{pilot}$ is the variance estimator assuming $\gamma_i$ are i.i.d. Gaussian variables.
When $a_{n}=o_{p}(n^{1/4})$, the penalty function in (\ref{eq:penalty_function}) satisfies the assumptions for our asymptotic theory.  A similar requirement on $a_n$ is made by \cite{Chen2009}.

\subsection{EM algorithm with Gauss-Hermite quadrature}
We propose an EM algorithm to maximize the penalized likelihood. At the $t$th iteration of the algorithm, given the parameter value $\bdtheta^{(t-1)}$ from the previous iteration, we first evaluate the following loss function at the E-step
\begin{eqnarray}\label{eq:regularEM_Q}
	Q(\bdtheta|\bdtheta^{(t-1)}) 
	& = & \sum_{i=1}^n E\left[\ell_{i, comp}(\theta; \BY_i , \BX_i, \gamma_i, \BL_i)|\BY_i,\BX_i,\bdtheta^{(t-1)}\right]  +\sum_{c=1}^Cp_n(\sigma^{2}_{c};\wh\sigma^2_{pilot}) 
\end{eqnarray}
where 
\bse
   	&&\hskip-10mm E\left[\ell_{i, comp}(\theta; \BY_i , \BX_i, \gamma_i, \BL_i)|\BY_i,\BX_i,\bdtheta^{(t-1)}\right]   \\
 	&& \hskip2cm = \sum_{c=1}^C \int \log f(\BY_{i}|\BX_{i},\gamma; \bdtheta_y) f(\gamma, L_{ic}=1|\BX_{i}, \BY_{i}; \bdtheta^{(t-1)}) d\gamma  \\
	&& \hskip2.5cm +\sum_{c=1}^C\int\left[\log f_c(\gamma| \mu_c, \sigma_c)
	f(\gamma, L_{ic}=1|\BX_{i}, \BY_{i}; \bdtheta^{(t-1)})\right]d\gamma  \\
  	&& \hskip2.5cm+\sum_{c=1}^C\log\pi_c\int f(\gamma, L_{ic}=1|\BX_{i}, \BY_{i}; \bdtheta^{(t-1)})d\gamma, \\
	&&\hskip-10mm f(\gamma, L_{ic}=1|\BX_{i}, \BY_{i}; \bdtheta^{(t-1)}) = \frac{\pi_c^{(t-1)}f(\BY_{i}|\BX_{i},\gamma; \bdtheta_y^{(t-1)}) {1 \over \sigma_c^{(t-1)} } \phi\left({ \gamma-\mu_c^{(t-1)} \over \sigma_c^{(t-1)}} \right)  } 
	{\sum_{c=1}^C \pi_{c}^{(t-1)} \int f(\BY_{i}|\BX_{i},\gamma; \bdtheta_y^{(t-1)}) {1\over \sigma_c^{(t-1)}} \phi\left({\gamma-\mu_c^{(t-1)} \over \sigma_c^{(t-1)} } \right) d\gamma}.
	%
	%
	%
\ese

Integrals with respect to a Gaussian density can be well approximated by Gauss-Hermite quadrature:
\bse
	 \int h(\gamma) \frac{1}{\sigma} \phi\{ (\gamma-\mu)/\sigma\} d\gamma \approx  \frac{1}{\sqrt{\pi}}\sum_{m=1}^{M}w_{m} h(\gamma_m)
\ese
where    $h(\gamma)$ is an integrable real valued function,  $\gamma_m = \mu+ \sqrt{2}\sigma d_m$, $d_1, d_2, \ldots, d_M$ are the Gauss-Hermite abscissas and $w_1, w_2, \ldots, w_M$  are the corresponding quadrature weights. We find in our numerical studies that using $M=100$ quadrature points usually provides a close enough approximation. More details on the Gauss-Hermite approximation of the loss function, $\wh Q(\bdtheta| \bdtheta^{(t-1)}) $, are provided in the supplementary material.

In the $M$-step, we maximize  $\wh Q(\bdtheta| \bdtheta^{(t-1)}) $ with respect to $\bdtheta$.  Define $\gamma^{(c,m)} = \mu_c^{(t-1)} + \sqrt{2}\sigma_c^{(t-1)} d_m$, 
\ben\label{eq:posterior_weights}
	\omega_{icm} = \frac{\tilde{\omega}_{icm} }{\sum_{c=1}^C \sum_{m=1}^M \tilde{\omega}_{icm} }, \quad \hbox{where } \tilde{\omega}_{icm} = w_{m}f(\BY_{i}|\BX_{i},\gamma^{(c,m)}; \bdtheta_y^{(t-1)} )\pi_{c}^{(t-1)}.
\een
We then update different components of $\bdtheta$ 
\begin{eqnarray*}
	&& {\pi}_{c}^{(t)}= \frac{1}{n} \sum_{i=1}^n\sum_{m=1}^{M} \omega_{icm},  \quad \quad 
	{\mu}_{c}^{(t)}=\frac{\sum_{i=1}^n \sum_{m=1}^{M}\gamma^{(c,m)} \omega_{icm} }
	{\sum_{i=1}^n\sum_{m=1}^{M}\omega_{icm}}, \\
	&& ({\sigma}_{c}^{2})^{(t)}=\frac{\sum_{i=1}^n\sum_{m=1}^{M}(\gamma^{(c,m)}-{\mu}^{(t)}_{c})^{2} \omega_{icm} +2a_{n}\wh{\sigma}_{pilot}^{2}}{\sum_{i=1}^n\sum_{m=1}^{M}\omega_{icm}+2a_{n}},
\end{eqnarray*}
and obtain $\bdtheta_y^{(t)}$ by maximizing
\[
\sum_{i=1}^n\sum_{c=1}^C\sum_{m=1}^{M} \omega_{icm} \log f(\BY_{i}|\BX_{i},\gamma^{(c,m)}; \bdtheta_y)
\]
using iteratively reweighted least squares.
We adopt the rule of \citet{Booth1999} and declare the algorithm converges at iteration $t$ if 
\[
\max_{l}\frac{|\theta_{l}^{(t)}-\theta_{l}^{(t-1)}|}{|\theta_{l}^{(t-1)}|+0.001}<0.001,
\]
where $\theta_l$ is the $l$th entry in $\bdtheta$.

At convergence, the weight $\omega_{icm}$ can be used to calculate some other quantities of interest, such as the marginal likelihood, the posterior probability of $\gamma_i$ belonging to the $c$th component and posterior mean of $\gamma_i$ . For example, we predict $\gamma_i$ by its posterior mean
\bse
	\int \gamma f(\gamma | \BY_i, \BX_i, \bdtheta)d\gamma= {\sum_{c=1}^C \pi_{c} \int \gamma f(\BY_{i}|\BX_{i},\gamma; \bdtheta_y )\phi\{(\gamma-\mu_c)/\sigma_c\}/\sigma_c d\gamma \over 
	\sum_{c=1}^C \pi_{c} \int f(\BY_{i}|\BX_{i},\gamma; \bdtheta_y)\phi\{(\gamma-\mu_c)/\sigma_c\}/\sigma_c d\gamma  }.
\ese
Using the Gauss-Hermite approximation, the posterior mean is approximated as
\ben\label{eq:posterior_mean}
\wh{\gamma}_i = \sum_{c=1}^C\sum_{m=1}^{M}\gamma^{(c,m)} \omega_{icm}
\een
where $\omega_{icm}$ is defined in (\ref{eq:posterior_weights}) evaluated at $\wh\bdtheta$. 

To obtain some reasonable initial values for $\bdtheta_y$ and $\bdtheta_{\gamma}$, we first run a generalized linear mixed model assuming $\gamma_i$'s are i.i.d. normal. We use the estimated fixed effects as initial values for $\boldsymbol{\theta}_y$, fit a Gaussian mixture model on the predicted values $\wh{\boldsymbol{\gamma}}$ and use the results as the initial values for $\bdtheta_\gamma$.

\subsection{Consistency of the estimator}\label{sec:consistency}
The EM algorithm essentially maximizes the following penalized marginal likelihood 
\begin{equation}\label{eq:pen_marg_like}
		l_{pen}(\bdtheta; \BY, \BX)= l_n(\bdtheta;\BY, \BX)+\sum_{c=1}^C p_n(\sigma_c^2),
\end{equation}
where
\ben\label{eq:marg_like}
	l_n(\bdtheta;\BY, \BX)=\sum_{i=1}^n \log\boldsymbol{\int} \left\{\prod_{k=1}^{N_i} f(Y_{ik}|\BX_{ik},\gamma; \bdtheta_y) g(\gamma|\bdtheta_{\gamma})\right\}d {\gamma}.
\een

The parameter space for a model with exactly $C$ components is
\ben\label{eq:parameter_space}
	\Theta_C=&\{ \bdtheta \mid \boldsymbol{\beta}\in		\mathbb{R}{}^{p}, \  
 		\ \mu_1< \mu_2 <\cdots < \mu_C,
		 \  \hbox{$\sum_{c=1}^{C}$}\pi_{c}=1, \\
& \ 0 < \pi_{c} < 1,
		\  \sigma_{c} > 0, 
		\ c=1,2,\ldots,C\}. \nonumber 
\een
The closure of $\Theta_C$  is 
$
	\bar{\Theta}_C=\{ \bdtheta \mid \boldsymbol{\beta}\in\mathbb{R}{}^{p}, 
	\ \hbox{$\sum_{c=1}^{C}$}\pi_{c}=1,
	\ 0 \le \pi_{c}\le 1, \ \mu_1\le \mu_2 \le \cdots \le \mu_C,\ \sigma_{c} \ge 0, 
	\ c=1,2,\ldots,C\},
$
which also includes the over-fitted models.
In other words, $\bar\Theta_C$ admits models where the true number of components is strictly less than $C$. There are multiple ways to parameterize an extra component in $\bar\Theta_C$. For example, setting either $\pi_c=0$ or $(\mu_c,\sigma_c) =(\mu_{c'}, \sigma_{c'})$ for some $c' \ne c$ means component $c$ does not exist. Various parameter values under these circumstances are identified as a single value, because they lead to the same mixture model. Let $\bdtheta_0\in \bar\Theta_C$ be the true parameter, $f(\bdx, \bdy | \bdtheta)$ be the joint distribution function of $(\BX, \BY)$ associated with the likelihood in (\ref{eq:marg_like}) and 
\ben\label{eq:equiv_param}
	\ \ \CF=\bigg\{\bdtheta\in \bar\Theta_C; \ \int_{-\infty}^{(\bdx', \bdy')} f(\bdx, \bdy | \bdtheta)d\mu(\bdx,\bdy) = \int^{(\bdx', \bdy')}_{-\infty} f(\bdx, \bdy, | \bdtheta_0)d\mu(\bdx,\bdy) \hbox { for any } (\bdx',\bdy')\bigg\}.
\een
Following \cite{Hathaway1985a}, we identify $\CF$ as a single point,  stated as Assumption 4 in the supplementary material.

Denote the maximum penalized likelihood estimator under a $C$-component mixture model by $
\wh{\bdtheta}_{C}=\arg\max_{\bdtheta\in \bar{\Theta}_{C}}l_{pen}(\bdtheta).
$
Because $\wh \bdtheta_C$ can be considered as a modified maximum likelihood estimator \citep{Kiefer1956},  its consistency follows from similar arguments as in \cite{Kiefer1956} and \cite{Hathaway1985a}.
The consistency for $\wh \bdtheta_C$ is established in the following proposition, the proof of which is relegated to the supplementary material.

\begin{pp}
\label{prop:consistency}
Under Assumptions 1-6 in the supplementary material, $\wh{\bdtheta}_C$ is consistent in the sense $\inf_{\bdtheta^\ast\in \CF} \|\wh\bdtheta_C- \bdtheta^\ast\|\to 0$ in probability.
\end{pp}


%
%
%
%
%
%
%
%
%
%
%
%
%
%
%
%
%
%
%
%
%
%
%
%
%
%
%
%

\section{Deciding the Number of Mixture Components\label{sec:NUMBER-OF-COMPONENTS}}

Deciding the number of components is key in answering whether there are subgroups of transplant centers that are under-performing or out-performing the rest. There are two commonly used approaches, the model selection approach \citep{ishwaran2011bayesian, Woo2006} and the hypothesis testing approach,  with different focuses as argued in \cite{Chen2012}. The model selection approach seeks a model to adequately describe the data, while the hypothesis testing approach is used to validate scientific claims. In this paper, we focus on the hypothesis testing approach because it quantifies the uncertainty of our decisions by providing $p$-values. Among many hypotheses that we can test, the most important one is $H_0: \ C_0=1$ vs $H_1: \ C_0=2$, where $C_0$ is the true number of components. This test is also referred to as the homogeneity test, since the null hypothesis means all transplant centers are from the same homogeneous population and none are under or over performing. If $H_0: \ C_0=1$ is rejected, we will also sequentially test other hypotheses of the form $H_0:\ C_0=C$ vs  $H_1: \ C_0=C+1$, $C=2,3,\ldots$, in search for the true number of components.

Because of the loss of strong identifiability for finite Gaussian mixture models, the regular asymptotic theory for likelihood ratio tests (LRT) does not hold. Instead, \citet{Chen2012} and \citet{Kasahara2014} proposed a locally restricted likelihood ratio test that confines the parameter space in a local alternative model to ensure the existence of an asymptotic distribution for the test statistic.  We extend such a test to the GLMM setting.

\subsection{Homogeneity Test}\label{sec:homogeneity_test}
We first consider  $H_0: C_0=1$ vs $H_1: C_0=2$. We  refer to the model under the null hypothesis as the reduced model and that under the alternative as the full model. When the null hypothesis is true, $\gamma_i$ are i.i.d. random variables following $\Normal(\mu_\gamma, \sigma_\gamma^2)$.  However, this model is not uniquely parameterized in the full model, unless we restrict the values of some parameters. Following \citet{Chen2012},  we restrict the parameter space under the full model to $\bar \Theta_2(\tau)=\{\bdtheta=(\mu_1,\mu_2,\sigma_1,\sigma_2, \pi_1,\pi_2)\trans; \ \mu_1,\mu_2\in \mathbb{R}, \sigma_1,\sigma_2 \ge 0, \pi_1=\tau, \pi_2=1-\tau\}$, for a fixed $\tau\in (0,0.5]$.  By doing so, we do not impose any constraints on the order between $\mu_1$ and $\mu_2$.
In $\bar\Theta_2(\tau)$, the null model is uniquely parameterized by $\bdtheta_0(\tau)=\{\bdtheta_{y,0}\trans, \bdtheta_{\gamma,0}\trans(\tau)\}\trans$, where $\bdtheta_{\gamma, 0}(\tau)=(\mu_\gamma, \mu_\gamma, \sigma_\gamma, \sigma_\gamma, \tau, 1-\tau)\trans$.

\subsubsection{Asymptotic Behavior of the Estimators}
Let $\bar\Theta_1$ be the parameter space when $C_0=1$ and the reduced model estimator be $\wh{\bdtheta}_{red}=\arg\max_{\bdtheta\in \bar{\Theta}_{1}} l_{pen}(\bdtheta)$, which is the usual MLE for GLMM under Gaussian random effect assumption. 
Under the full model, the estimator under a fixed $\tau$ is
$$
	\wh{\bdtheta}_{full}(\tau)=\arg \max _{\bdtheta\in \bar\Theta_{2}(\tau)} l_{pen} (\bdtheta).
$$ 
This estimator can be obtained using the EM algorithm described in Section \ref{estimation} without the step for updating $\pi_c$'s. The following proposition provides the convergence rate of $\wh{\bdtheta}_{full}(\tau)$ under the null hypothesis. 

\begin{pp}
\label{pp:convergencerate1}
Under $H_0: C_0=1$ and Assumptions 1-7 in the supplementary material, for any fixed $\tau\in (0,0.5]$,  $\wh{\boldsymbol{\beta}}_{full}(\tau) - \boldsymbol{\beta}_0 = O_p(n^{-1/2})$, and $\wh \bdtheta_{\gamma, full}(\tau)-\bdtheta_{\gamma, 0}(\tau)=O_p(n^{-1/4})$. 
\end{pp}

\noindent{\bf Remark:} We use a similar reparameterization as \citet{Kasahara2014} in the proof of Proposition \ref{pp:convergencerate1}. As shown in the proof, many derivatives of the log likelihood are either exactly zero or have mean zero, and it takes a ninth order Taylor expansion to get a local quadratic approximation to the penalized likelihood. The convergence rate in the proposition means that, for an over-fitted mixture model, the GLMM regression coefficient $\bdbeta$ still enjoys the root-$n$ convergence rate, while the parameters of the latent Gaussian mixture model converge much slower. This slow convergence rate also stresses a fundamental difference between our latent Gaussian mixture model and the common parametric models. 

\subsubsection{Test Procedure}
Let $\BCT$ be any subset of numbers in $(0,0.5]$, define the test statistic 
\begin{equation}\label{eq:teststat1}
	\wt T_{1}=\max_{\tau\in\boldsymbol{\mathcal{T}}} T_1(\tau) \quad 
	\hbox{where  } T_1(\tau)=2[ l_n\{\wh{\bdtheta}_{full}(\tau)\}-l_n(\wh{\bdtheta}_{red})].
\end{equation}

\begin{pp}
\label{prop:T1distn}
Under $H_0: C_0=1$ and Assumptions 1-7,  $\wt T_{1}\cid \chi^{2}(2)$ as $n\to\infty$. 
\end{pp}

\noindent{\bf Remark:} Our proof of Proposition \ref{prop:T1distn} shows that, under $H_0: C_0=1$, $T_1(\tau) \cid \chi^2(2)$ for any fixed $\tau$. In fact, if there is only one true component, no matter how we choose to split that component, the leading term in the asymptotic expansion of $T_1(\tau)$ remains the same. We define $\wt T_1$ as the maximum of $T_1(\tau)$ over $\CT$ to increase the power: if $H_1$ is true, the more values of $\tau$ we try, the better chance we have to detect an extra component. Proposition \ref{prop:T1distn} holds if $\wt T_1$ is the maximum of $T_1(\tau)$ over the whole interval $(0,0.5]$, but for practical consideration $\CT$ is often taken as a finite subset.

The detailed test procedure is given as follows.

\noindent{\bf Step 0.} Obtain $\wh \bdtheta_{red}$ and $l_n(\wh\bdtheta_{red})$.

\noindent{\bf Step 1.} For a fixed $\tau$, obtain $\wh \bdtheta_{full}(\tau)$. To guarantee a global maximum of the penalized likelihood is reached, try 100 randomly selected initial values for $\bdtheta(\tau)$.

\noindent{\bf Step 2.} (Optional) Using $\wh{\bdtheta}_{full}(\tau)$ obtained in Step 1 as the starting value,
perform two more EM iterations without fixing $\tau$, and use the resulting estimator to evaluate $T_1(\tau)$. 

\noindent{\bf Step 3.} Repeat Steps 1 and 2 for each $\tau\in \boldsymbol{\mathcal{T}}$ to obtain $\wt T_1$, where $\boldsymbol{\mathcal{T}}$ is set to be $\{0.1,0.3,0.5\}$ following the recommendation of \cite{Chen2012}. 

\noindent{\bf Step 4.} For a size $\alpha$ test, reject $H_0: C_0=1$ if $\wt T_1 > \chi^2_\alpha(2)$.

In Step 2, we perform two more EM iterations without fixing $\tau$ to increase the power of the test, which is the recommendation of \cite{Chen2012}.





%
%
%
%
%
%
%
%
%
%
%
%
%
%
%
%
%

\subsection{Testing for C greater than 2} \label{sec:high_order_test}

Next, we consider a test  $H_0: C_0=C$ vs $H_1: C_0=C+1$ for a $C\ge2$. We now refer to the model with $C$ components as the reduced model and that with $C+1$ components as the full model.  
We first estimate the reduced model and let the reduced model estimator be $\wh{\bdtheta}_{red}=\arg\max_{\bdtheta\in \bar{\Theta}_{C}} l_{pen}(\bdtheta)$. Assuming $H_0$ is true, denote the true value of the parameter by $\bdtheta_0$ and order the true mean parameters by $\mu_{1,0}<\mu_{2,0}<\cdots <\mu_{C,0}$. This parameter is not uniquely identified in the full model: if any $\pi_c=0$ or $(\mu_c,\sigma_c)=(\mu_{c+1},\sigma_{c+1})$ for some $c\in \{1,2,\ldots, C\}$, the full model degenerates to the reduced model. In order to make the reduced model identifiable in $\bar\Theta_{C+1}$, we will impose constraints that $\pi_c>0$ for all $c=1,\ldots, C+1$ and $\pi_c/(\pi_c+\pi_{c+1})=\tau$ for some $c$ and a fixed $\tau\in (0,0.5]$ like we did in Section \ref{sec:homogeneity_test}.

\subsubsection{Locally Restricted Full Model Estimators}
To test if a $(C+1)$-component mixture model fits the data better, we will test to see if any one of the $C$ components in the reduced model can be further split into two.  Define non-overlapping intervals $D_1, \ldots, D_C$ such that $\mu_{c,0}\in D_c$. For a fixed $\tau \in (0,0.5]$ and $c\in \{1,\ldots, C\}$, define neighborhoods in the parameter space $\bar\Theta_{C+1}$
\bse
	 \CN_{C+1}(c, \tau)&=&\{ \bdtheta \in \bar \Theta_{C+1} \mid 	
	\hbox{$\frac{\pi_{c}}{\pi_{c}+\pi_{c+1}}=\tau$};  \quad \mu_{c'}\in D_{c'} \hbox{ for $c'<c$}; \\
	&&\hskip20mm \hbox{$ \mu_c, \mu_{c+1}\in D_c$;  \quad $\mu_{c'}\in D_{c'-1}$ for $c'>c+1$} \}.
%
\ese
The neighborhood $\CN_{C+1}(c,\tau)$ collects the parameters that split the $c$th component into two daughter components with a split proportion $\tau$, while restricting the other mean parameters from changing too much. The definition of $\CN_{C+1}(c,\tau)$ requires knowledge about intervals $\{D_1,D_2,\ldots, D_C\}$ that contain the true mean parameters. In practice, we already have consistent estimator of $\mu_{c,0}$ from fitting the reduced model, replacing $\{D_c\}_{c=1}^C$ with their consistent estimates does not affect the asymptotic behavior of the test we are about to propose. A practical choice for $\{D_c\}_{c=1}^C$ is provided below in the test procedure. Like in Section \ref{sec:homogeneity_test}, we do not restrict order between $\mu_c$ and $\mu_{c+1}$ in $\CN_{C+1}(c,\tau)$ because $\tau$ is restricted in $(0,0.5]$.

Define the locally restricted full model estimator as
$$
	\wh{\bdtheta}_{full}(c, \tau)=\arg \max _{\bdtheta\in \CN_{C+1}(c,\tau)} l_{pen} (\bdtheta).
$$ 
To obtain this estimator, we need some minor adjustments to the EM algorithm in Section \ref{estimation}. First, we update $\pi_c+\pi_{c+1}$ as a single parameter and then assign values for $\pi_c$ and $\pi_{c+1}$ proportional to $\tau$. Second, after each $M$-step, we enforce the restrictions in $\CN_{C+1}(c,\tau)$ by forcing any $\mu_{c'}$ stepping out of boundary back to its predetermined range. A similar scheme is used in \cite{Chen2012}.


The following convergence rate result echoes Proposition \ref{pp:convergencerate1}. It shows that the component that we are trying to split suffers a slower convergence rate, because it is overfitted in $\CN_{C+1}(c,\tau)$ as a mixture of two daughter components, and the rest of the parameters converge in root-$n$ rate.
\begin{pp}
\label{pp:convergencerateC}
Under $H_0: C_0=C$ and Assumptions 1-8 in the supplementary material, for any fixed $\tau\in (0,0.5]$, then
\bse
	&&\wh \mu_{c,full}(c,\tau)-\mu_{c,0}=O_p(n^{-1/4}), 
	\quad \wh \mu_{c+1,full}(c,\tau)-\mu_{c,0}=O_p(n^{-1/4}), \\
	&&\wh \sigma_{c,full}(c,\tau)-\sigma_{c,0}=O_p(n^{-1/4}),
	\quad \wh \sigma_{c+1,full}(c,\tau)-\sigma_{c,0}=O_p(n^{-1/4}),
\ese
and 
$
\wh \bdtheta_{y,full}(c,\tau)-\bdtheta_{y 0} = O_p(n^{-1/2})$, 
$\wh \bdtheta_{\gamma, c',full}(c,\tau)-\bdtheta_{\gamma,c',0}= O_p(n^{-1/2})$ for $c'< c$, 
$\wh \bdtheta_{\gamma, c',full}(c,\tau)-\bdtheta_{\gamma, c'-1,0}= O_p(n^{-1/2})$ for  $c'> c+1$, where $\bdtheta_{\gamma, c'}=(\mu_{c'}, \sigma_{c'}, \pi_{c'})\trans$.
\end{pp}

\subsubsection{Local Reparameterization, Test Statistic and Asymptotics}

To test if any component in the reduced model can be further divided into two, define the test statistic
\ben\label{eq:T_C_tau}
	T_C(\tau)= \max_{c\in \{1,2,\ldots,C\}} T_C(c,\tau), \quad \hbox{where } T_C(c,\tau)= 2[ l_n\{\wh\bdtheta_{full}(c,\tau)\}- l_n(\wh\bdtheta_{red})].
\een
Let $\CT$ be any finite subset of $(0,0.5]$, define test statistic
\ben\label{eq:T_C}
	\wt T_C=\max_{\tau\in \CT} T_C(\tau).
\een

In order to understand the asymptotic behavior of $T_C(c,\tau)$, we adopt the reparameterization of \cite{Kasahara2014} in $\CN_{C+1}(c,\tau)$. Define the new parameter vector as $\bdpsi(c,\tau)=(\bdtheta_y\trans, \boldsymbol{\delta}(c)\trans, \bdmu(c)\trans, \bdsigma^2(c)\trans, \lambda_{\mu},\lambda_{\sigma})\trans$ such that
 \ben\label{eq:reparam_C}
	\left(\begin{array}{c}
	\mu_{c}\\
	\mu_{c+1}\\
	\sigma_{c}^{2}\\
	\sigma_{c+1}^{2}
	\end{array}\right)=\left(\begin{array}{c}
	\nu_{\mu}+(1-\tau)\lambda_{\mu}\\
	\nu_{\mu}-\tau\lambda_{\mu}\\
	\nu_{\sigma}+(1-\tau)(2\lambda_{\sigma}-\frac{1+\tau}{3}\lambda_{\mu}^{2})\\
	\nu_{\sigma}-\tau(2\lambda_{\sigma}+\frac{2-\tau}{3}\lambda_{\mu}^{2})
	\end{array}\right),
\een
and
\ben\label{eq:reparam_C_rest}
\begin{array}{lllcl}
	 \boldsymbol{\delta}(c) &=& (\pi_1,  \ldots, \pi_{c-1}, &\pi_c+\pi_{c+1},& \pi_{c+2}, \ldots,\pi_{C})\trans, \\
	 \bdmu (c)&=&(\mu_1, \ldots, \mu_{c-1},&\nu_{\mu},& \mu_{c+2}, \ldots,  \mu_{C},\: \mu_{C+1})\trans, \\
	 \bdsigma^2(c) & =&(\sigma^2_1, \ldots,\sigma^2_{c-1}, &\nu_{\sigma},& \sigma^2_{c+2}, \ldots, \sigma^2_{C}, \:\sigma^2_{C+1})\trans.
 \end{array}
\een

Denote the new parameter space as $\bar{\Theta}_{\psi, C+1}$ and
partition $\bdpsi$ into $(\bdeta\trans,\boldsymbol{\lambda}\trans)\trans$
where 
\bse
	\bdeta&=&\{\bdtheta_y\trans, \boldsymbol{\delta}(c)\trans, \bdmu(c)\trans, \bdsigma^2(c)\trans\}\trans, \quad 
	\boldsymbol{\lambda}=(\lambda_{\mu},\lambda_{\sigma})\trans.
\ese
The reduced model is uniquely parameterized by $\bdtheta^*\in \CN_{C+1}(c,\tau)$, and it is reparameterized as $\bdpsi^{*}=\{(\bdeta^{*})\trans,0,0\}\trans$, or more specifically 
$\bdtheta_y=\bdtheta_{y,0}$, $\boldsymbol{\lambda}^*=\boldsymbol{0}$ and
$\boldsymbol{\delta}^*(c)= ( \pi_{1,0}, \pi_{2,0}, \ldots,\pi_{C-1,0} ) \trans$, $\bdmu^*(c)=(\mu_{1,0}, \mu_{2,0}, \ldots, \mu_{C,0})\trans$, $\bdsigma^{2*}(c) =(\sigma^{2}_{1,0},\sigma^{2}_{2,0},  \ldots, \sigma^{2}_{C,0})\trans$. The benefit of the reparameterization (\ref{eq:reparam_C}) is that, to test if the $c$th component can be further split, we can equivalently test if $\bdlambda={\pmb 0}$.

Define the score function with respect to $\bdpsi(c, \tau)$ as
\ben\label{eq:score_C}
	\bds^{(c)}_i=\left\{\bds_{\bdeta,i}\trans, (\bds^{(c)}_{\boldsymbol{\lambda},i})\trans \right\}\trans,
\een
where
 \begin{eqnarray*}
	 \bds_{\bdeta,i}&=&\left(\begin{array}{c}
			\bds_{\bdtheta_y,i}\\
			\bds_{\boldsymbol{\delta},i}\\
			\bds_{\mu, i}\\
			\bds_{\sigma, i}
			\end{array}\right), 
	\quad \quad
	\bds_{\lambda, i}^{(c)} = \left(\frac{\int\zeta_{i}\pi_{c}f_{c,i}^{*}H_{ci}^{3*}}{\int\zeta_{i}g^{*}},\frac{\int\zeta_{i}\pi_{c}f_{c,i}^{*}H_{ci}^{4*}}{\int\zeta_{i}g^{*}}\right)\trans,\\
	\bds_{\bdtheta_y, i}&=&\frac{\int (\partial \zeta_{i}/\partial \bdtheta_y) g^{*}}{\int\zeta_{i}g^{*}}, \\
	 \bds_{\boldsymbol{\delta},i}&=&\left(\frac{\int\zeta_{i}(f_{1,i}^{*}-f_{C,i}^{*})}{\int\zeta_{i}g_{i}^{*}},\ldots \frac{\int\zeta_{i}(f_{C-1,i}^{*}-f_{C,i}^{*})}{\int\zeta_{i}g_{i}^{*}}\right)\trans, \\
	\bds_{\mu,i}&=&\left(\frac{\int\zeta_{i}\pi_{1}f_{1,i}^{*}H_{1i}^{1*}}{\int\zeta_{i}g^{*}},\ldots,\frac{\int\zeta_{i}\pi_{C}f_{C,i}^{*}H_{Ci}^{1*}}{\int\zeta_{i}g^{*}}\right)\trans, \\
	\bds_{\sigma,i}&=&\left(\frac{\int\zeta_{i}\pi_{1}f_{1,i}^{*}H_{1i}^{2*}}{\int\zeta_{i}g^{*}},\ldots,\frac{\int\zeta_{i}\pi_{C}f_{C,i}^{*}H_{Ci}^{2*}}{\int\zeta_{i}g^{*}}\right)\trans.
\end{eqnarray*}
Here, we use the short hand notation $\zeta_{i} = \prod_{k=1}^{N_i} f(y_{ik}|\bdx_{ik},\gamma_{i};	\bdtheta_y)$, $f_{c,i}^{*}=f_c(\gamma_{i}|\mu_{c,0}, \sigma_{c,0})$,
$g_{i}^{*}=g(\gamma_{i}|\bdtheta_{\gamma}^{*})$ and $H_{ci}^{k*}=H^k\left(\frac{\gamma_i -\mu_{c,0}}{\sigma_{c,0}}\right)/(k!\sigma_{c,0}^{k})$, where $H^k(\cdot)$ is the $k$th Hermite Polynomial.


\begin{pp}
\label{prop:TCdistn}

Under $H_0: C_0=C$ and Assumptions 1-8 in the supplementary material, 
\bse
	\wt T_{C}\cid
	\max\left\{(\boldsymbol{S}_{\lambda|\eta, n}^{(c)})^{T}(\boldsymbol{{\cal I}}_{\lambda|\eta}^{(c)})^{-1} \boldsymbol{S}_{\lambda|\eta, n}^{(c)},c=1,2,\ldots,C\right\},
\ese
where 
$\boldsymbol{S}_{\lambda|\eta, n}^{(c)} = \boldsymbol{S}_{\lambda, n}^{(c)}-\boldsymbol{\mathcal{I}}^{(c)}_{\lambda\eta}\boldsymbol{\mathcal{I}}_{\eta}^{-1}\boldsymbol{S}_{\eta, n}$, 
$\boldsymbol{{\cal I}}^{(c)}_{\lambda|\eta}=
\boldsymbol{{\cal I}}^{(c)}_{\lambda}-\boldsymbol{{\cal I}}^{(c)}_{\lambda\eta}\boldsymbol{{\cal I}}_{\eta}^{-1}(\boldsymbol{{\cal I}}^{(c)}_{\lambda\eta})^{T}$, 
$\boldsymbol{S}_{\eta, n}=\frac{1}{\sqrt{n}}\sum_{i=1}^{n}\bds_{\eta, i}$, $\boldsymbol{S}_{\lambda, n}^{(c)}=\frac{1}{\sqrt{n}}\sum_{i=1}^{n}\bds_{\lambda, i}^{(c)}$,
$\boldsymbol{{\cal I}}^{(c)}_{\lambda\eta} = E\{\bds_{\boldsymbol{\lambda},i}^{(c)} \bds_{\bdeta, i}\trans\}$, 
$\BCI_\eta=E( \bds_{\eta, n} \bds_{\eta, n}\trans)$, and
$\boldsymbol{{\cal I}}^{(c)}_{\lambda} = E\{\bds_{\boldsymbol{\lambda},i}^{(c)} (\bds_{\boldsymbol{\lambda}, i}^{(c)})\trans\}$
\end{pp}

One can show $(\boldsymbol{S}_{\lambda|\eta ,n}^{(c)})^{T}(\boldsymbol{{\cal I}}_{\lambda|\eta}^{(c)})^{-1} \boldsymbol{S}_{\lambda|\eta, n}^{(c)} \cid \chi^2(2)$ for each $c$, but the score vectors $\boldsymbol{S}_{\lambda|\eta, n}^{(c)}$ are dependent among different $c$'s and hence the distribution of $\wt T_{C}$ in Proposition \ref{prop:TCdistn} is that of the maximum of a few correlated $\chi^2(2)$ random variables. In Appendix A, we describe a simulation method to evaluate this asymptotic distribution. This procedure only requires estimation of the covariance matrix of $\{ \BS_{\lambda|\eta,n}^{(c)}, c=1,\ldots, C\}$ and simulating Gaussian random variables. It is extremely fast and fundamentally different from bootstrap, which requires fitting the model a large number of times to the bootstrap samples.

%
%
%
%
%
%
%
%
%
\subsubsection{Test Procedure}
For any $C\ge 2$, our test procedure for $H_0: C_0=C$ is as follows.

\noindent{\bf Step 0.} Obtain $\wh \bdtheta_{red}$ using penalty function (\ref{eq:penalty_function}) and $a_n=\frac{1}{n}$, and evaluate $l_n(\wh\bdtheta_{red})$.
Define subintervals $D_1=[\wh{\gamma}_{min},\frac{\wh{\mu}_{1,red}+\wh{\mu}_{2,red}}{2}]$, $D_2=(\frac{\wh{\mu}_{1,red}+\wh{\mu}_{2,red}}{2},\frac{\wh{\mu}_{3,red}+\wh{\mu}_{2,red}}{2}]$, $\ldots D_C=(\frac{\wh{\mu}_{C-1,red}+\wh{\mu}_{C,red}}{2}, \wh{\gamma}_{max}]$, where $\wh{\gamma}_{min}$ and $\wh{\gamma}_{max}$ are the minimum and maximum of the predicted ${\gamma}$'s.

\noindent{\bf Step 1.} Obtain $\wh{\bdtheta}_{full}(c, \tau)$ by maximizing the penalized likelihood in the restricted parameter neighborhood $\CN_{C+1}(c,\tau)$ using the subintervals $\{D_k\}_{k=1}^C$ defined in Step 0. The penalty on $\sigma_{k}^2$ is $p_n(\sigma_{k}^2, \wh{\sigma}^2_{c',red})$ if $\mu_k $ is restricted in $D_{c'}$, $k=1,\ldots, C+1$, and $a_{n}$ is chosen according equation (23) in \citet{Kasahara2014}. If a $\mu_k$ steps outside of its range $D_{c'}$ specified by $\CN_{C+1}(c,\tau)$ during the EM iterations, we will simply set it back to the nearest boundary of $D_{c'}$. To ensure that the maximum of $l_{pen}$ is reached, we repeat the EM algorithm 100 times using randomly selected initial values within $\CN_{C+1}(c,\tau)$. 


\noindent{\bf Step 2.} Using $\wh{\bdtheta}_{full}(c, \tau)$ as starting value,
do two more EM iterations without fixing $\tau$. Use the resulted estimator to evaluate $T_C(c,\tau)$ in (\ref{eq:T_C_tau}).


\noindent{\bf Step 3.} Repeat Steps 1 and 2 for each $c=1, 2,\ldots, C$,  and for each $\tau \in \CT=\{0.1,0.3, 0.5\}$, and evaluate $\wt T_C$ in (\ref{eq:T_C}).

\noindent{\bf Step 4.}
Evaluate the asymptotic null distribution in Proposition \ref{prop:TCdistn} using the procedure described in Appendix A and compare $\wt T_{C}$ with the null distribution to get the $p$ value.

\subsection{Sequential Test to Determine the Order of the Latent Gaussian Mixture Model}
Hypothesis tests are not designed for model selection, but can nevertheless be used for such a purpose in an exploratory study. One can determine the order of  the latent Gaussian mixture model by sequentially testing $H_{01}: C_0=1$, $H_{02}: C_0=2$, $H_{03}: C_0=3$, $\ldots$, and declare $C_0=C^\ast$ if $H_{0 C^\ast}$ is the first null hypothesis in the sequence that is not rejected. Such a procedure is obviously not a consistent model selection procedure, as we have a fixed chance to fail to reject a hypothesis. On the other hand, one can also argue many widely used model selection procedures are not consistent, such as the Akaike Information Criterion. To control the family wise error rate at $\alpha$, one can adopt a Bonferroni procedure and set the sizes of the tests to be $\alpha/2$, $\alpha/4$, $\alpha/8$, $\ldots$. 

%
%
%
%
%
%
%
%
%
%
%
%
%
%
%
%
%

\section{Transplant Center Evaluation with False Discovery Rate Control}\label{sec:fdr}
One important goal of our study is to provide a ranking for the transplant centers. The evaluation is based on the value of the latent variable $\gamma$, which represents the care quality of a center. For methodology development, we first assume that the number of mixture components $C_0$ is correctly specified and all parameters in the latent Gaussian mixture model are known. 

Following \cite{Efron2004}, we identify the ``empirical null'' distribution  of $\gamma$ as a subset of components in the mixture density, $g_0(\gamma| \bdtheta_\gamma)=\sum_{c\in \CC_0}\pi_c f_c(\gamma| \mu_c, \sigma_c)/ \sum_{c \in \CC_0 } \pi_c$ where $\CC_0\subset\{1,2,\ldots, C\}$. For each transplant center $i$, we will test if this center belongs to one of the components in $\CC_0$, or $H_{i0}: \sum_{c\in \CC_0} L_{ic}=1$, $i=1,\ldots, n$. Suppose $\CC_0$ consists of centers of average performance, then center $i$ is considered ``interesting'' (either outperforming or underperforming) if $H_{i0}$ is rejected.

Since $\gamma_i$ is not directly observed, our decision rule for $H_{i0}$  is based on the observed data $\BX_i$ and $\BY_i$, denoted as $\delta_i=\delta(\BX_i,\BY_i; \bdtheta)$, where $\delta_i=1$ means center $i$ is ``interesting'' and $\delta_i=0$ otherwise. The false discovery rate is defined as  
\bse
	FDR = E\left\{\frac{\sum_i^n I(\delta_i=1,\sum_{c\in \CC_0}L_{ic}=1) }{\sum_i^n I(\delta_i=1)}  \Big| \sum_i^n I(\delta_i=1) > 0\right\} P\left\{\sum_i^n I(\delta_i=1) > 0\right\}
\ese
When $\gamma_i$'s are observed, \cite{Sun2007} show that the oracle decision rule is based on the local FDR, $T_{\rm OR}(\gamma_i)=P(\sum_{c\in \CC_0} L_{ic}=1 | \gamma_i)=\sum_{c \in \CC_0 } \pi_c f_c(\gamma_i) / g(\gamma_i)$. 
In our case, $\gamma_i$ is not observed, and the local FDR is defined as 
\ben\label{eq:lfdr}
	lFDR_i&=&P(\hbox{$\sum_{c\in \CC_0}L_{ic}=1 $} | \BX_i, \BY_i) \nonumber \\	
	&=& \frac{\left(\sum_{c \in \CC_0 } \pi_c\right) \int f(\BY_i | \BX_i, \gamma; \bdbeta) g_0(\gamma| \bdtheta_{\gamma}) d\gamma} 
	{\int f(\BY_i | \BX_i, \gamma; \bdbeta) g(\gamma | \bdtheta_{\gamma}) d\gamma} \nonumber \\
	&=&\frac{\sum_{c\in \mathcal{C}_0} \pi_{c} \int f(\BY_i | \BX_i, \gamma; \bdbeta) f_c(\gamma | \mu_c,\sigma_c) d\gamma  
	 }
	{\int f(\BY_i | \BX_i, \gamma; \bdbeta) g(\gamma | \bdtheta_{\gamma})d\gamma }.
\een
It is easy to show $lFDR_i= E\{ T_{\rm OR}(\gamma_i) | \BX_i, \BY_i\} $. Following \cite{sun2015false}, the multiple hypotheses testing problem is related to a classification problem with the loss function
\bse\label{eq:loss} 
	\MSL(\BL, \bddelta)=\hbox{$\lambda \sum_i \delta_i (\sum_{c\in \CC_0}L_{ic}) + \sum_{i} (1-\delta_i)(1-\sum_{c\in \CC_0}L_{ic} $}),
\ese
where $\lambda$ is a penalty for false positive. Let $\MSR=E \{ \MSL(\BL, \bddelta)\}$ be the risk of the classification problem, and by Theorem 1 of \cite{sun2015false}, the optimal decision rule that minimizes this risk is $\delta_i= I(lFDR_i< t)$ for some threshold $t$.

Let ${lFDR}_{(1)} \le {lFDR}_{(2)} \le \cdots \le {lFDR}_{(n)}$ be the ranked lFDR values. For any $\alpha>0$, let $k=\max_{i} \{{1\over i} \sum_{j=1}^i lFDR_{(j)} \le \alpha\}$ and our FDR control procedure is to reject all $H_{i0}$ with the rank of $lFDR_{i}$ less or equal to $k$. 


%
%
\begin{pp}
\label{prop:FDR}
Under the model in (\ref{eq:glmm}), the above procedure controls FDR at level $\alpha$.
\end{pp}
A sketch proof of Proposition \ref{prop:FDR} is provided in Section \ref{sec:proof_fdr} of the supplementary material. 
In practice, $lFDR$ is estimated by substituting $\bdtheta$ with its estimator and the integrals in (\ref{eq:lfdr}) are evaluated using Gaussian quadrature as described above.

\section{Simulation Studies}\label{sec:simulations}

We conduct simulation studies to examine the numerical performance of proposed estimation procedure and the validity and power of the proposed tests in choosing the order of the latent Gaussian mixture model.

\subsection{Simulation 1: Estimation and Random Effect Prediction}\label{sec:simu_est}

We simulate data for $n=282$ transplant centers, which is the number of kidney transplant centers in OPTN in year 2008. The number of patients per center has a highly skewed distribution in the real data. To mimic such a distribution, we generate $N_i$ as the integer part of  the sum of $Poission(5)$ and $Exponential(45)$. The response $Y_{ik}$ is a binary variable generated using (\ref{eq:glmm}) with $P(Y_{ik}=1)=\{1+\exp(-\xi_{ik})\}^{-1}$, where $\xi_{ik}=\BX_{ik}\trans \bdbeta +\gamma_i$. $\BX$ is generated from bivariate standard normal and 
$\boldsymbol{\beta}=(1, 1)\trans$.
In the following subsections, we generate $\gamma_i$'s from Gaussian mixture models with different orders.

\subsubsection{Two-Component Model}
We first generate $\gamma_i$'s from a two-component Gaussian mixture model
 \bse\label{eq:simu_model1}
 	\hbox{ Model 1: } \quad \quad 0.5 \ \Normal(-3.26,1.2^2)+0.5 \ \Normal(0.74,0.8^2).
%
\ese
The parameters in Model 1 are selected such that the marginal probability of $\{ Y_{ik}=1\}$ is roughly the same as the real data. We repeat the simulation 200 times and apply the estimation procedure in Section \ref{estimation} to each simulated data set. The mixture components in the estimated model are ranked according to the value $\wh\mu_c$ to avoid the cluster label switching problem. The results for parameter estimation under correctly specified number of components are summarized in Table \ref{tab:simu_model1}. As we can see, all estimators perform well: the biases are much smaller than the standard deviations, showing that our estimators are asymptotically unbiased. 

\begin{table}[ht]
\centering
\begin{tabular}{rrrrr}
  \hline
 & Truth & Mean & Bias & Std \\ 
  \hline
$\pi_1$ & 0.5000 & 0.4971 & -0.0029 & 0.0280 \\ 
  $\pi_2$ & 0.5000 & 0.5029 & 0.0029 & 0.0280 \\ 
  $\mu_1$ & -3.2598 & -3.2586 & 0.0012 & 0.1262 \\ 
  $\mu_2$ & 0.7402 & 0.7401 & -0.0001 & 0.0752 \\ 
  $\sigma_1$ & 1.2000 & 1.1954 & -0.0046 & 0.1340 \\ 
  $\sigma_2$ & 0.8000 & 0.7960 & -0.0040 & 0.0630 \\ 
  $\beta_1$ & 1.0000 & 1.0017 & 0.0017 & 0.0213 \\ 
  $\beta_2$ & 1.0000 & 1.0006 & 0.0006 & 0.0225 \\ 
   \hline
\end{tabular}
\caption{Summary for parameter estimation under Simulation Model 1 based on 200 replications.\label{tab:simu_model1}} 
\end{table}

\begin{figure}[H]
\begin{centering}
\includegraphics[scale=0.4]{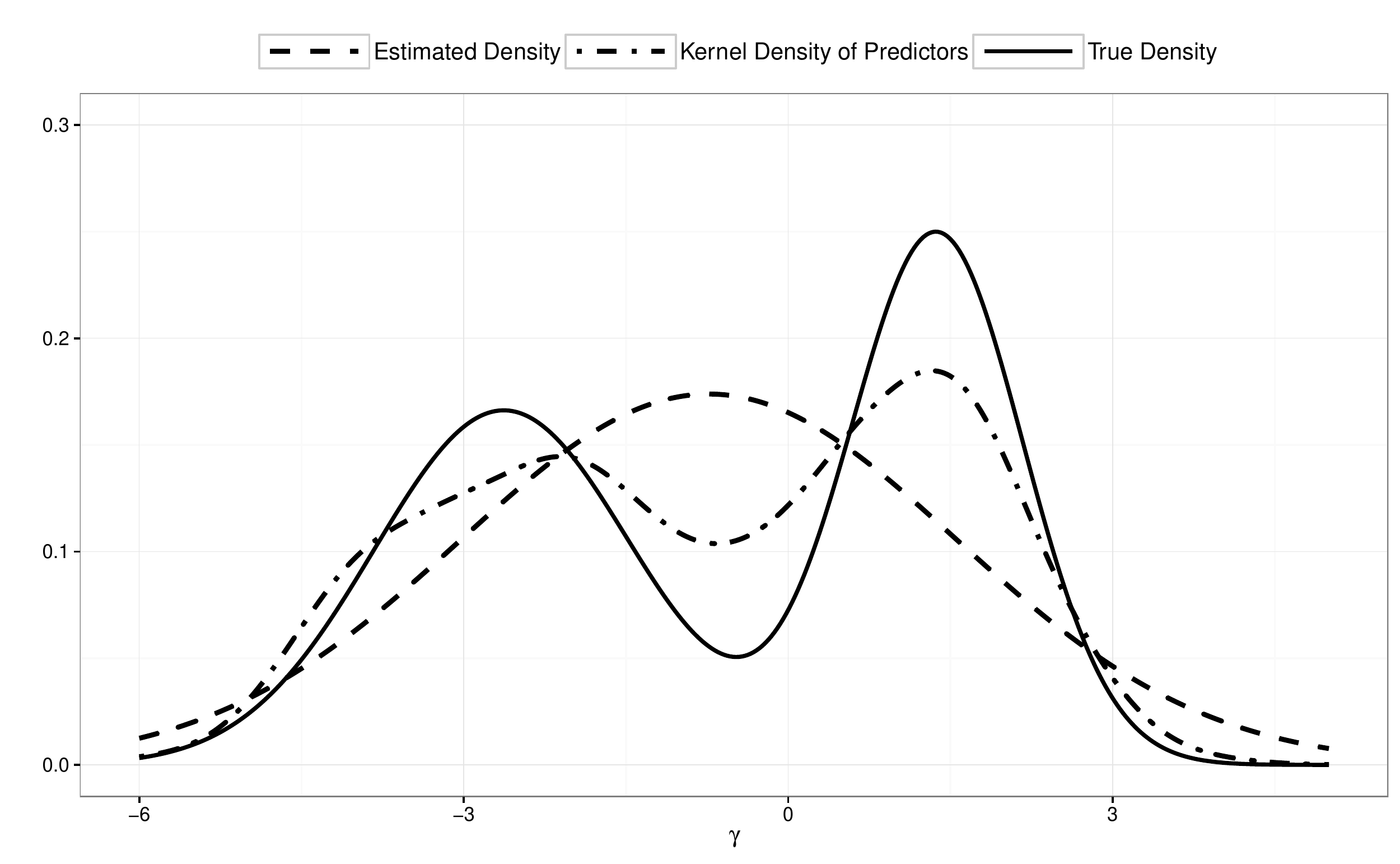}
\includegraphics[scale=0.4]{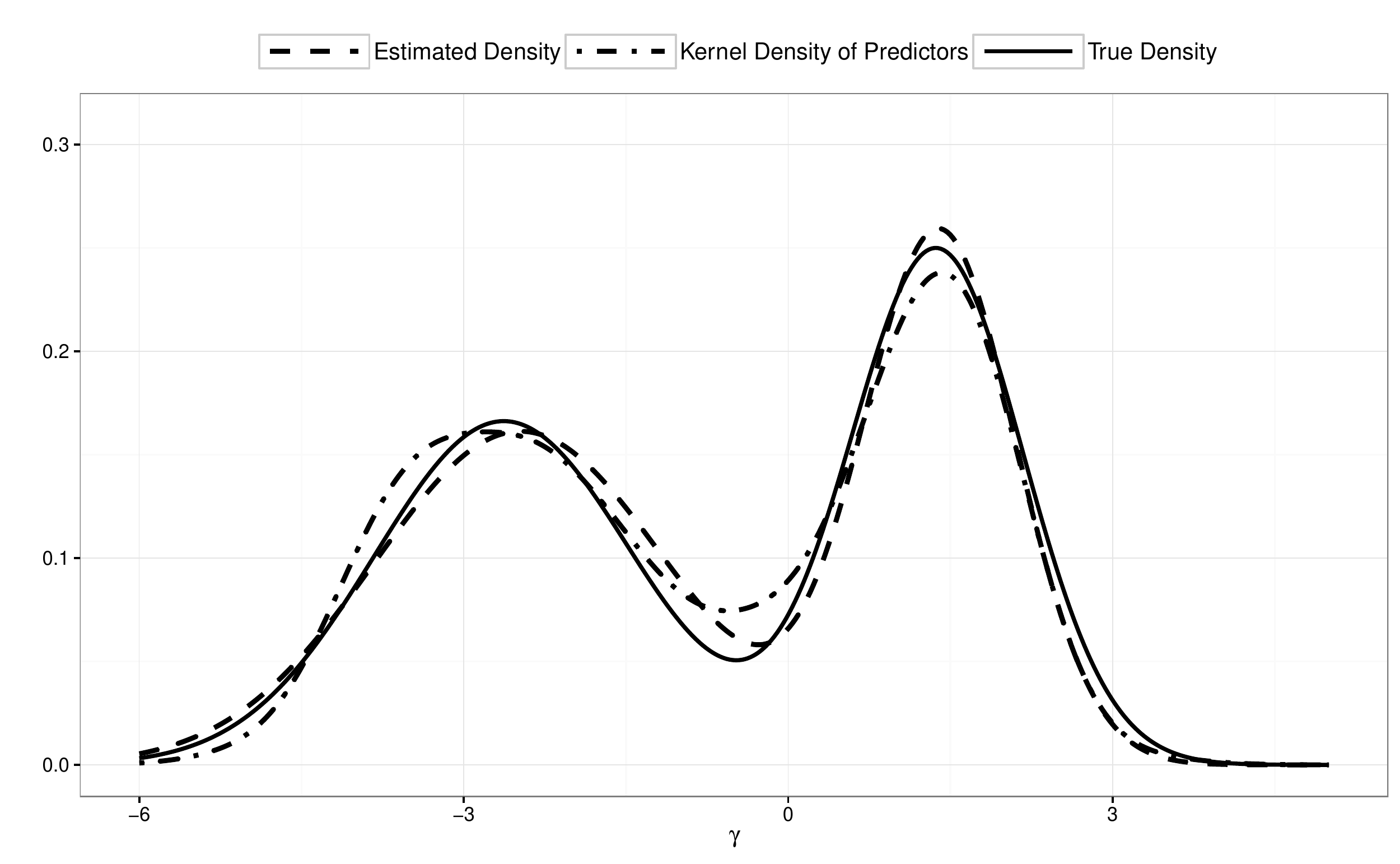}
\par\end{centering}
\caption{Simulation Model 1: impact of random effect assumption. Top panel: results from a common GLMM model with a mis-specified Gaussian random effect assumption; bottom panel: results of the proposed GLMM with latent Gaussian mixture random effects under correctly specified number of components. In both panels, the solid curve is the true density for $\gamma$, the dashed curve is the estimated density of $\gamma$ using the fitted model, and the dot-dash curve is the kernel density of the predicted random effects.  \label{toy_example}}
\end{figure}

To illustrate the drawback for mis-specifying the random effect distribution, we also fit a common GLMM model to the simulated data under the assumption that $\gamma_i$'s are i.i.d. Gaussian. Figure \ref{toy_example} illustrates the results in a typical simulation run. The upper panel shows the results of a common GLMM, and the lower panel shows the results of the proposed model. In both panels, we compare the true density of $\gamma$ with the estimated density using the fitted model and the kernel density of the predicted $\gamma$ using the fitted model. As we can see from the upper panel, prediction under the mis-specified Gaussian random effect assumption suffers from a shrinkage effect that the values of $\wh\gamma$ are pushed towards the center of the distribution so that the posterior distribution resembles the shape of a Gaussian distribution. The lower panel shows that prediction under our proposed model does not suffer from such a shrinkage effect. Our model recovers the shape of the latent variable distribution and produces better predictions. In Table \ref{tab:mspe}, we also report the mean square prediction error for the random effect averaged over the 200 simulation runs and the Monte Carlo standard deviation of the prediction error. As we can see, when the random effect distribution is mis-specified as Gaussian, the fitted model yields a much larger prediction error.

\subsubsection{Three-Component Model}

We repeat the simulation study while generating $\gamma_i$'s from the following three-component Gaussian mixture model 
\bse\label{eq:simu_model2}
	\hbox{Model 2:} \quad  0.3 \ \Normal(-5.26, 1.2^2) + 0.4 \ \Normal( -0.26, 0.8^2) + 0.3 \ \Normal(2.74, 0.9^2).
%
\ese
We repeat the simulation 200 times, perform the proposed estimation procedure under correctly specified order of mixture, and the estimation results are summarized in Table \ref{tab:simu_model2}. We can see that the estimation results are quite reasonable: all biases are virtually zero; the standard errors for component means ($\mu_c$) and component standard deviations ($\sigma_c$) are slightly inflated compared with Table \ref{tab:simu_model1}, which is understandable since we are fitting a more complicated mixture model; the standard errors for $\bdbeta$ are not affected by the increased complicity of the latent mixture model.

\begin{table}[ht]
\centering
\begin{tabular}{rrrrr}
  \hline
 & Truth & Mean & Bias & Std \\ 
  \hline
$\pi_1$ & 0.3000 & 0.3016 & 0.0016 & 0.0244 \\ 
  $\pi_2$ & 0.4000 & 0.3904 & -0.0096 & 0.0588 \\ 
  $\pi_3$ & 0.3000 & 0.3080 & 0.0080 & 0.0596 \\ 
  $\mu_1$ & -5.2598 & -5.2800 & -0.0202 & 0.2175 \\ 
  $\mu_2$ & -0.2598 & -0.2652 & -0.0054 & 0.3472 \\ 
  $\mu_3$ & 2.7402 & 2.6894 & -0.0508 & 0.3433 \\ 
  $\sigma_1$ & 1.2000 & 1.1821 & -0.0179 & 0.2664 \\ 
  $\sigma_2$ & 0.8000 & 0.8036 & 0.0036 & 0.1948 \\ 
  $\sigma_3$ & 0.9000 & 0.9286 & 0.0286 & 0.2516 \\ 
  $\beta_1$ & 1.0000 & 1.0010 & 0.0010 & 0.0225 \\ 
  $\beta_2$ & 1.0000 & 1.0038 & 0.0038 & 0.0226 \\ 
   \hline
\end{tabular}
\caption{Summary for parameter estimation under Simulation Model 2 based on 200 replications. \label{tab:simu_model2}} 
\end{table}



In Table \ref{tab:mspe}, we also present the mean square prediction error of the proposed model averaged over 200 simulation runs, Monte Carlo standard deviation of the prediction error, and the same quantities under GLMM with Gaussian random effects. As we can see the prediction error under the common GLMM with Gaussian assumption has much bigger prediction error than the proposed model. The gap between the prediction errors from the two models is even bigger than for Model 1, because Model 2 is even more heterogeneous.

\begin{table}[ht]
\centering
\begin{tabular}{crrr}
  \hline
 Simulation Model&  Fitted Model & Mean & Std \\ 
  \hline
Model 1 & GLMM Gaussian & 0.4167 & 0.0392\\ 
		& GLMM Mixture	 &0.3589 & 0.0361\\
   \hline
Model 2 & GLMM Gaussian  & 0.6988 & 0.0697\\ 
		&GLMM Mixture &0.5405 & 0.0581 \\
 \hline\hline 
\end{tabular}
\caption{Mean squared prediction error for the random effect under Simulation Models 1 and 2. GLMM Gaussian: generalized linear mixed model with Gaussian random effects; GLMM Mixture: the proposed model; Mean: Mean Squared Prediction Error averaged over 200 replicates; Std: standard deviation of the prediction error.  \label{tab:mspe}} 
\end{table}



\subsection{Simulation 2: Hypothesis Tests}\label{sec:simu_test}
Next, we investigate the validity and power for the proposed tests in Section \ref{sec:NUMBER-OF-COMPONENTS}.

\subsubsection{Asymptotic Null Distributions}
\begin{figure}
\begin{centering}
\includegraphics[scale=0.4]{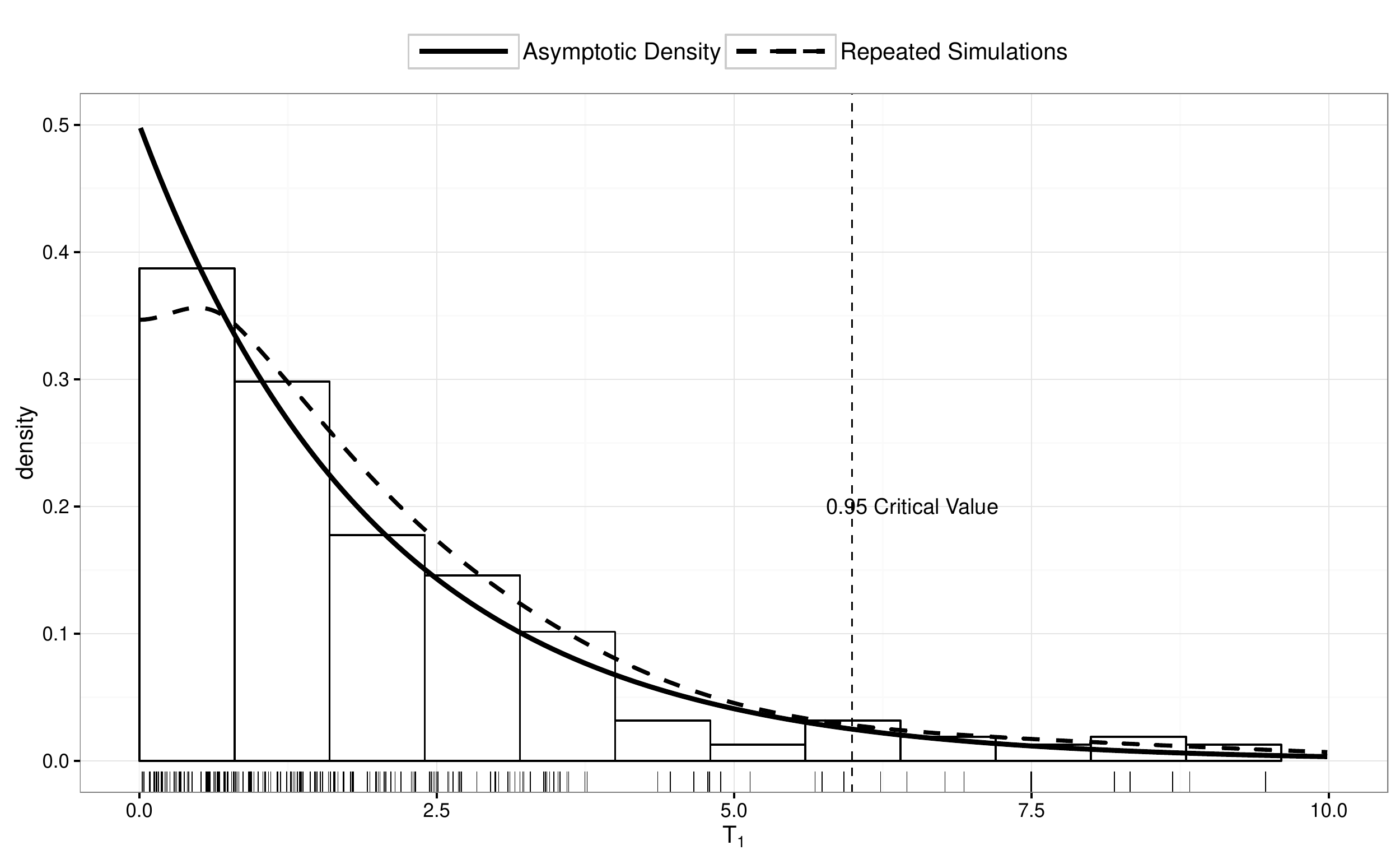}
\par\end{centering}
\begin{centering}
\includegraphics[scale=0.4]{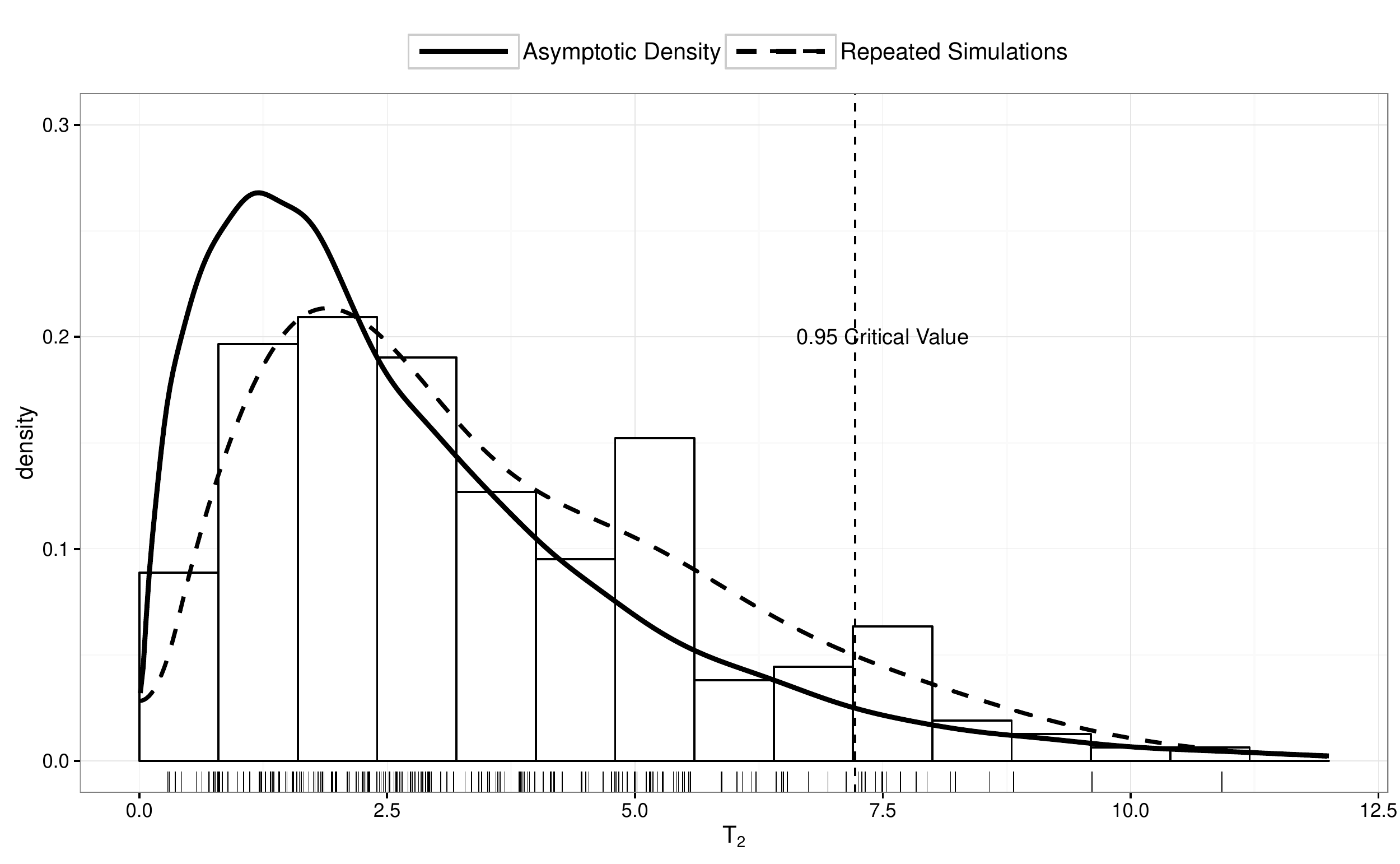}
\par\end{centering}
\begin{centering}
\includegraphics[scale=0.4]{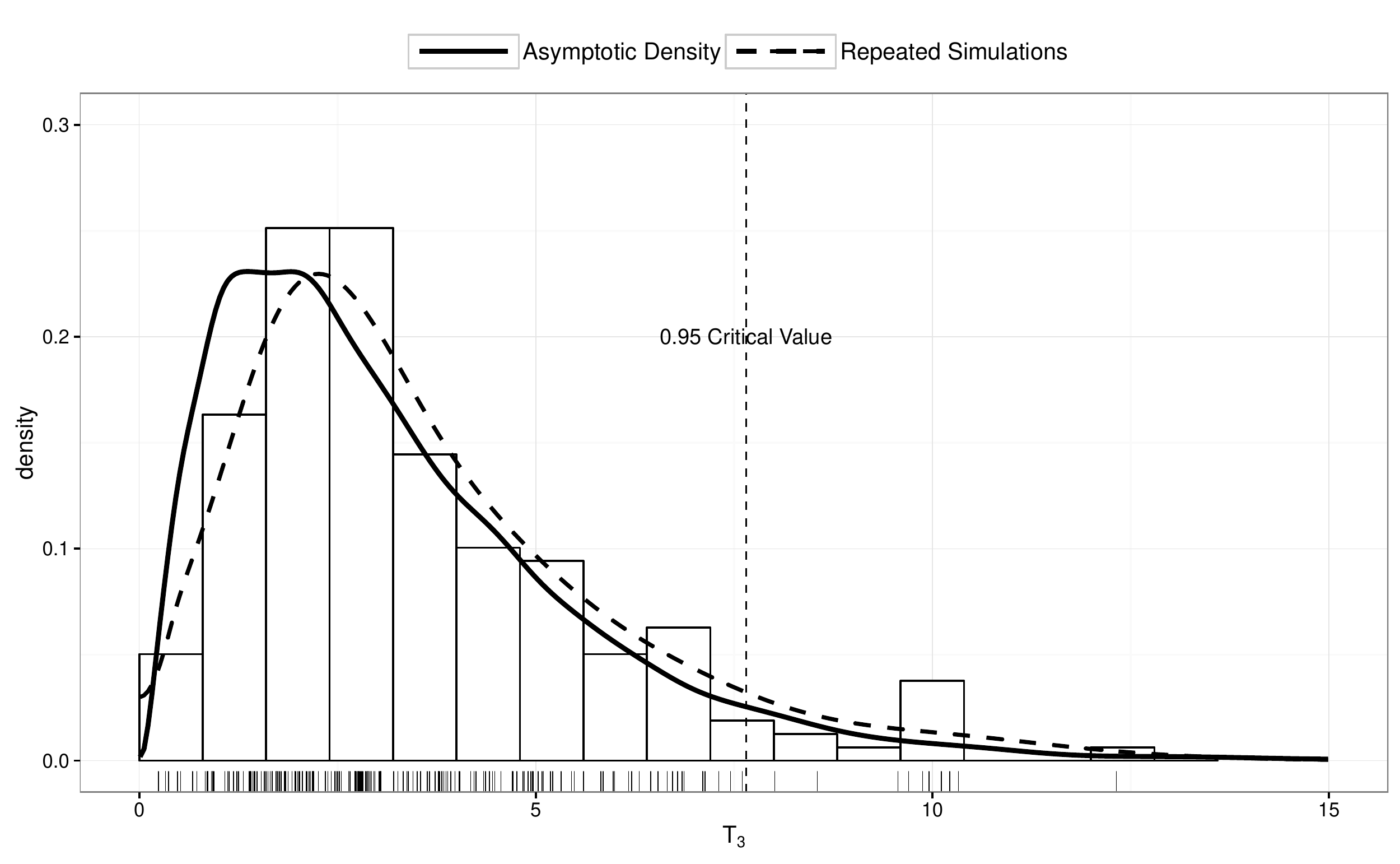}
\par\end{centering}
\caption{Empirical and asymptotic distributions of $T_{1}$, $T_{2}$ and $T_{3}$ under the null hypotheses.\label{fig:Asymptotic-distribution}}
\end{figure}

We generate simulated data under similar settings as in Simulation 1, while $\gamma_i$'s are generated from three models: Model 1, Model 2 and  
\bse\label{eq:simu_model0}
	\hbox{ Model 0:} \quad \quad \Normal(-1.26,0.5^2).
\ese
The three models represent latent Gaussian mixture models with orders 1 to 3. We generate 200 simulated data sets under each of the three models, and compute $\wt T_{1}$ in data under Model 0, $\wt T_{2}$ under Model 1 and $\wt T_{3}$ under Model 2. The empirical distributions of the three quantities represent the null distribution for the test statistics under the null hypotheses $C_0=1, 2$ and $3$ respectively. These empirical distributions are provided in Figure \ref{fig:Asymptotic-distribution} and compared with the asymptotic distributions provided in Section \ref{sec:NUMBER-OF-COMPONENTS}. In each panel of Figure \ref{fig:Asymptotic-distribution}, the dash curve is the kernel density based on 200 replicates of the test statistic and solid curve is the asymptotic distribution. Note that the asymptotic distribution for $\wt T_2$ and $\wt T_3$ are based on 10,000 simulations using the procedure described in Appendix A.  As we can see, the empirical distributions of the test statistics are remarkably close to the asymptotic distribution, which also shows the validity of the proposed tests.

\subsubsection{Power of the tests}\label{sec:simu_power}
Next, we illustrate the power of the tests. The response $Y$ is generated the same way as in Section \ref{sec:simu_est}, while $\gamma$ is generated from the following two models
\bse
	&& \hbox{Model 3:} \quad 0.6\Normal(-2.26,1.2^2)+0.4\Normal(-0.46,0.8^2), \\
	 && \hbox{Model 4:} \quad 0.3 \Normal(-3.26, 1.2^2) + 0.4 \ \Normal( -0.26, 0.8^2) + 0.3 \ \Normal(2.34, 0.9^2). 
\ese
Compared with Models 1 and 2 considered in Section \ref{sec:simu_est}, the individual components in Models 3 and 4 are less separated, making it harder to detect the real order of these models especially when $\gamma$ is an unobserved latent variable.

\begin{figure}
\begin{centering}
\includegraphics[scale=0.4]{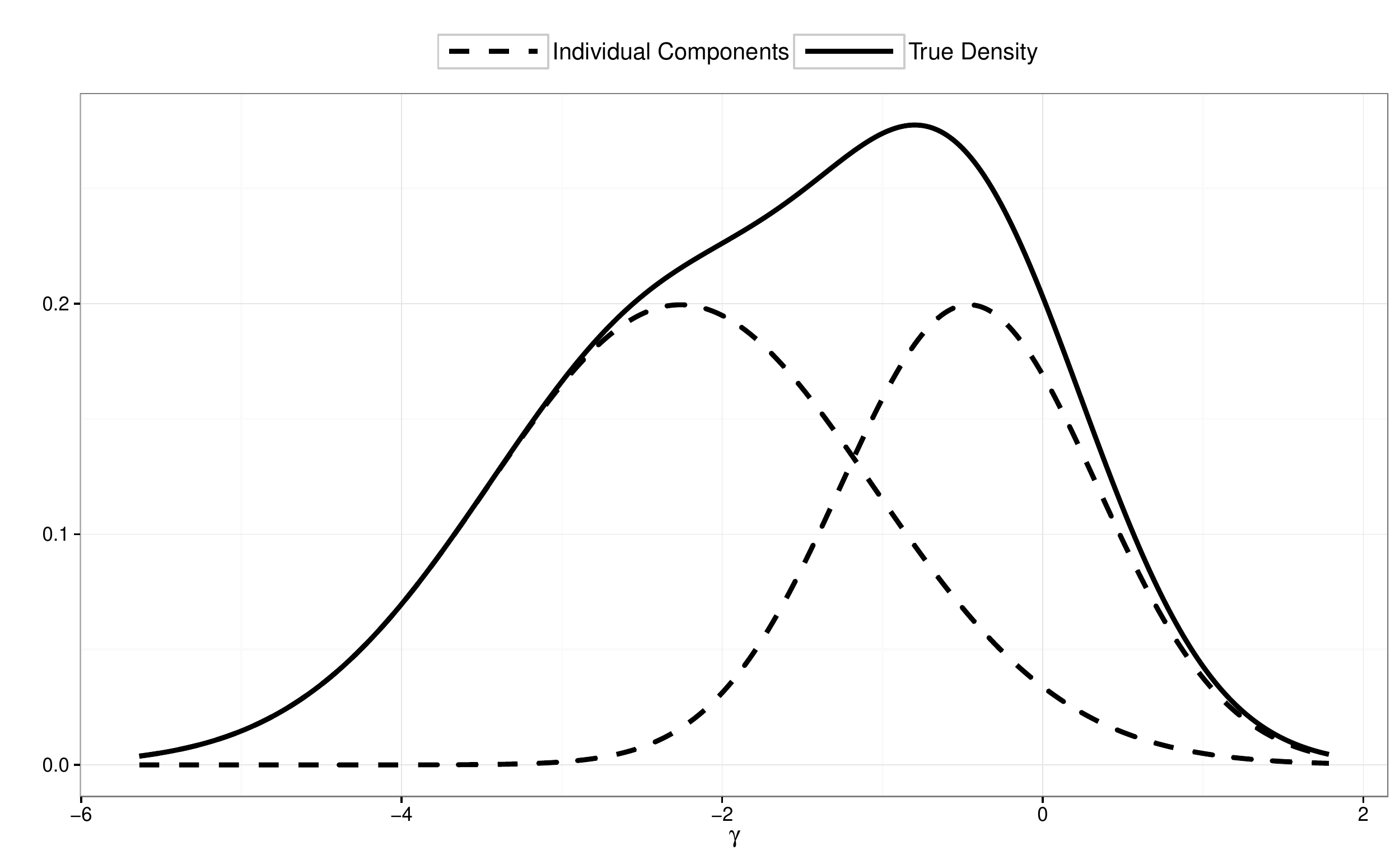}
\includegraphics[scale=0.4]{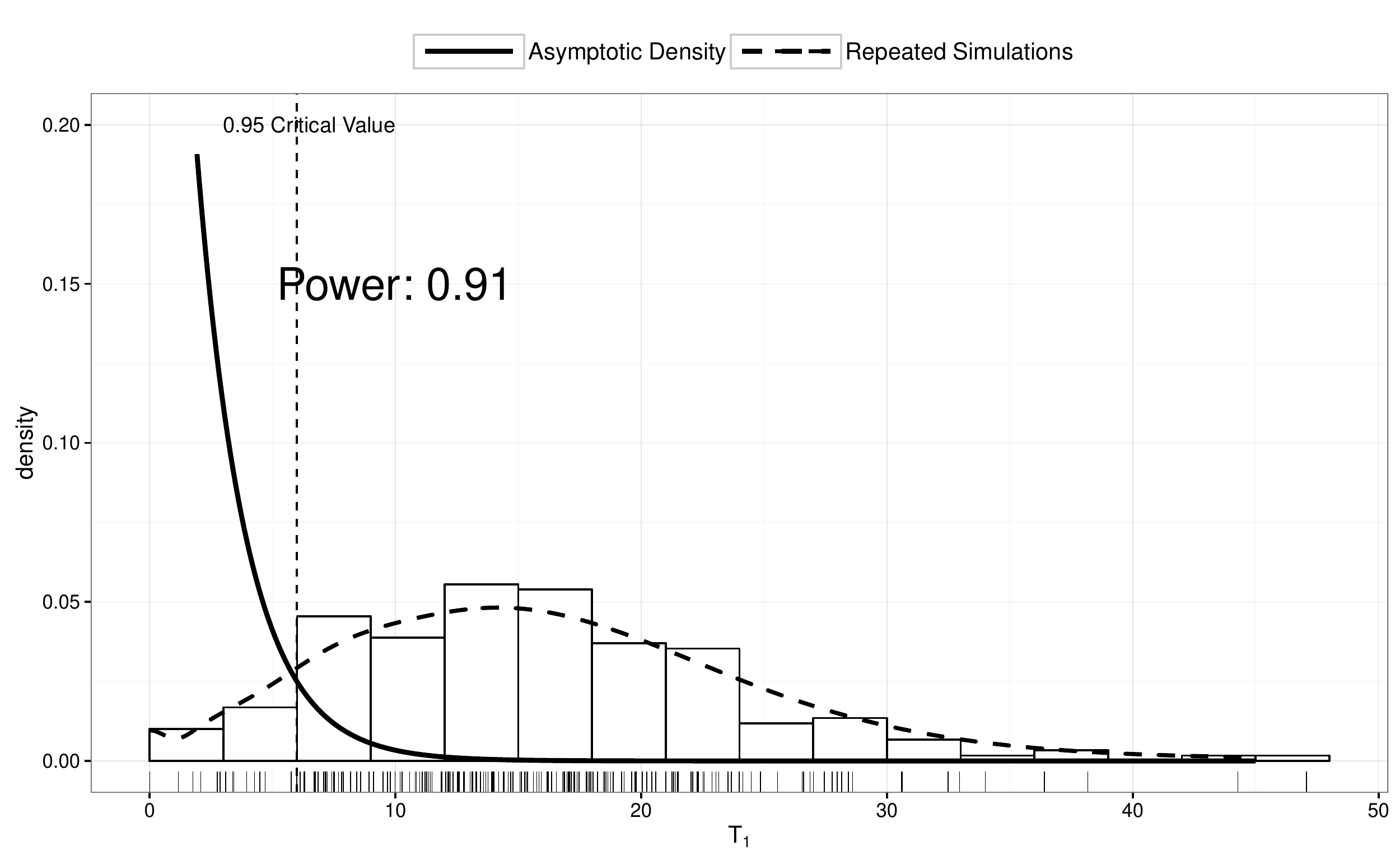}
\caption{Power of the homogeneity test. The top panel illustrates the true density of $\gamma$; the bottom panel shows the empirical distribution of $\wt T_1$ compared with the asymptotic null distribution.\label{fig:Power-of-thehomotest}}
\par\end{centering}
\end{figure}

To examine the power of the homogeneity test in Section \ref{sec:homogeneity_test}, we compute $\wt T_1$ in 200 simulated data sets where $\gamma_i$'s are simulated from Model 3, and summarize the results in Figure \ref{fig:Power-of-thehomotest}. In the top panel of Figure \ref{fig:Power-of-thehomotest}, we illustrate  the true density of $\gamma$ under Model 3; in the bottom panel, we compare the empirical distribution of $\wt T_1$ with its asymptotic distribution under $H_0: C_0=1$. If we perform a 5\% test based on the asymptotic $\chi^2(2)$ distribution, the power of the homogeneity test is 91\% under this scenario.

\begin{figure}
\begin{centering}
\includegraphics[scale=0.4]{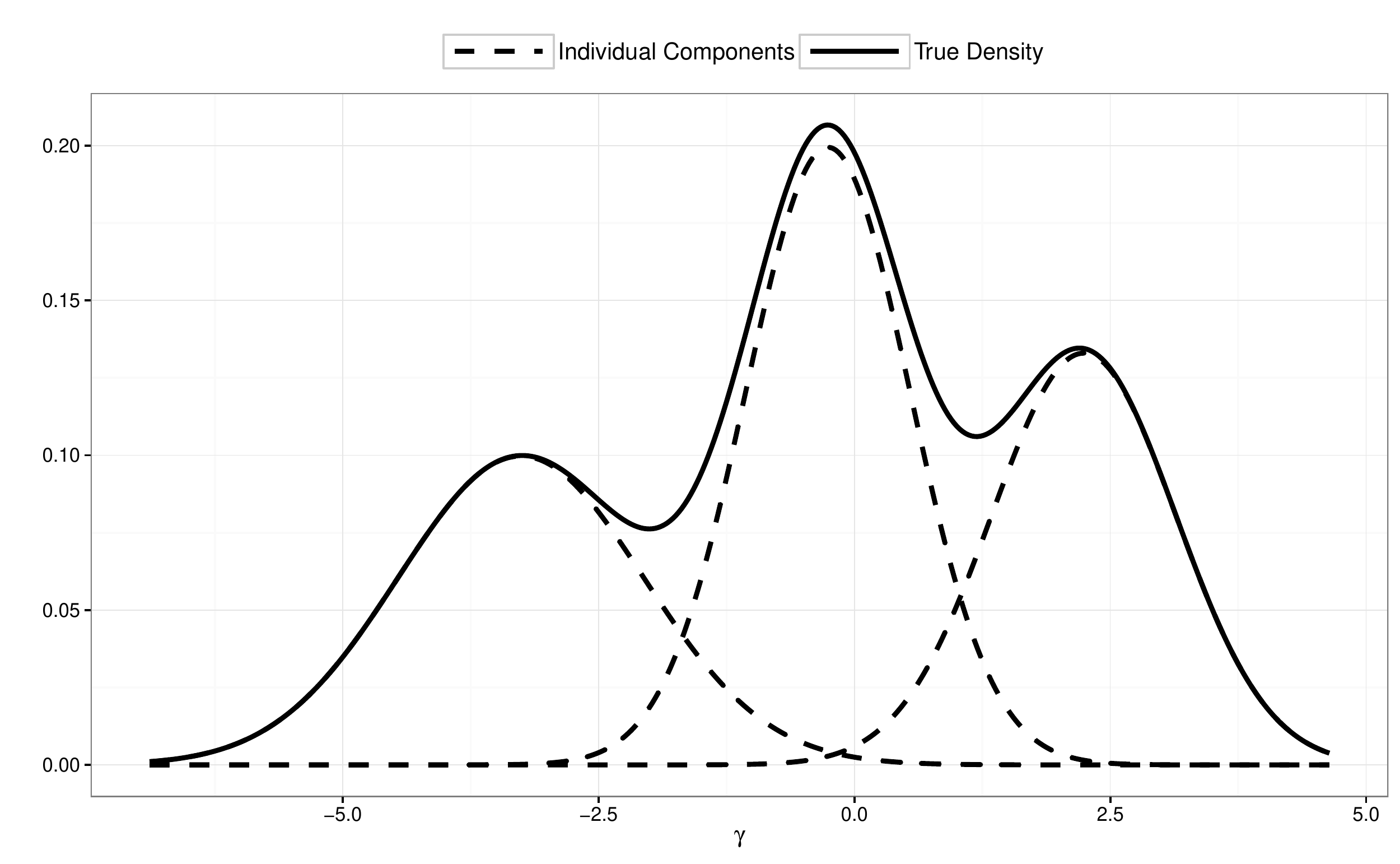}
\includegraphics[scale=0.4]{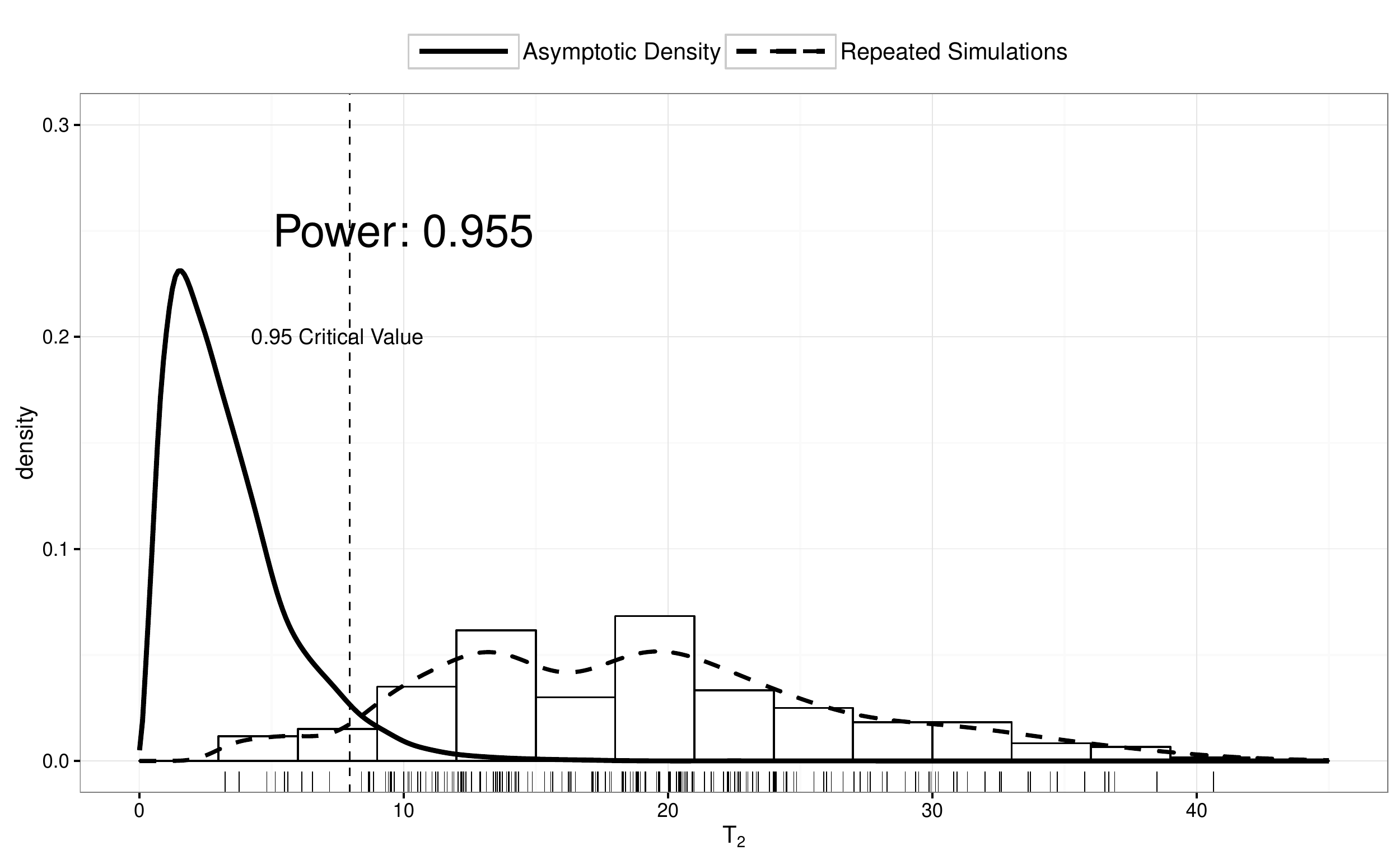}
\caption{Power of the locally restricted likelihood ratio test for $H_0: C_0=2$ vs $H_1: C_0=3$. The top panel illustrates the true density of $\gamma$; the bottom panel shows the empirical distribution of $\wt T_2$ compared with its asymptotic distribution under $H_0$. \label{fig:Power-of-theC2test}}
\par\end{centering}
\end{figure}

To examine the power of the locally restricted likelihood ratio test proposed in Section \ref{sec:high_order_test}, we perform test on $H_0: C_0=2$ vs $H_1: C_0=3$, while $\gamma_i$'s are simulated from Model 4. In Figure \ref{fig:Power-of-theC2test}, we illustrate the true density of $\gamma$ under Model 4, and compare the empirical distribution of $\wt T_2$ over 200 simulation runs with its asymptotic null distribution. The empirical power of the proposed test is 95.5\%.

We have also examined the power of the homogeneity test when $\gamma_i$'s are simulated from Model 1 and the power of the test on $H_0: C_0=2$ when $\gamma_i$'s are generated from Model 2. The power under both of these cases virtually equal to 1.

Since a sequential test can be used for model selection purpose, it is of interest to compare the test based procedure with other model selection procedures such as the Bayesian information criterion (BIC), which is the negative log likelihood for the observed data plus a penalty on $\log (n)$ times the number of free parameters in the model. We apply BIC to simulated data under both Model 3 and 4. For Model 3, BIC picks the correct model with 2 components in 39\% out of the 200 simulations and chooses a 1-component model for the remaining 61\% of the repetitions. This means if we use BIC as the decision rule to test $H_0: C_0=1$ under Model 3, it only has 39\% of power, which is much lower than the test we developed. 
For Model 4, BIC chooses a correct 3-component model in 50.5\% of the 200 simulations and chooses 1 or 2 components in the other 49.5\% of runs. On the other hand, the sequential test procedure with $\alpha=0.05$ chooses the correct number of components $88.5\%$ of the time for Model 3, and $86\%$ of the time for Model 4.  
 
%

 %
%
%
%
%
%
%
%
%
%
%
%
%
%
%
%
%
%

\section{Data Analysis}\label{sec:data}
\begin{figure}
\centering \includegraphics[scale=0.4]{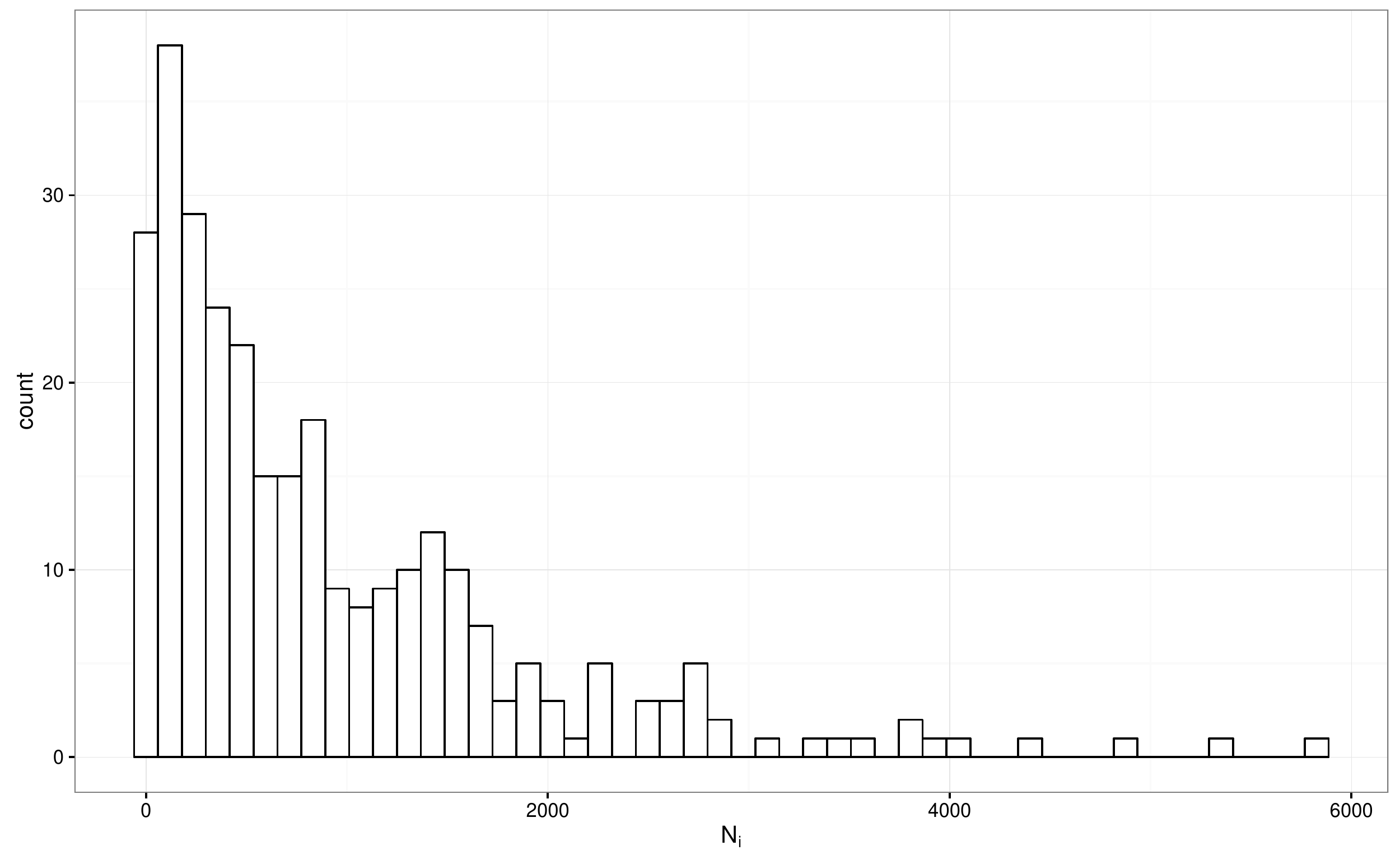}
\caption{Histogram for the number of patients per center in the OPTN data. 
\label{fig:Ni}}
\end{figure}

Our motivating data are obtained from the Organ Procurement and
Transplantation Network (OPTN),  administered  under a contract with the U.S. Department of
Health and Human Services (HHS). The OPTN data system includes data on all donor, wait-listed candidates, and transplant recipients in the US. Included in the analysis are adult renal
failure patients ($\geq18$ years of age) who underwent deceased donor kidney transplantation between January $1987$ and December $2008$. This cohort includes $N=269,386$ patients receiving kidney transplants from a total of $n=296$ centers. The number of transplants performed by a center, $N_i$, has a highly skewed distribution as illustrated in Figure \ref{fig:Ni}. Most centers performed a few hundred cases of kidney transplantation, but there are centers took over 5000 cases.
The patient level response is the 5-year survival status (1=death and -1=survival) and there is no censoring due to routine and rigorous tracking of the patients. The overall failure rate within 5 years of transplantation is $27.59\%$.

An important patient level covariate that is directly related to the success of kidney transplant is $x_1=$ cold ischemic time, which is the time that the donor kidney was kept in a refrigerator before received by the patient.
Other patient level covariates include  $x_2=$ age at transplantation and $x_3=$ sex of the patient (1 =male, 0=female), $x_4$ -- $x_6$ are indicators for BMI in the intervals  (22, 25], (25-30] and 30+ respectively. 
Since the data were collected in a time span of two decades, it is possible that the technology used in transplant surgeries has been improving over time which also affects the patient level outcome. Therefore, we also include time effects into the model in additional to the other covariates described above. Using cases before 1990 as baseline, covariates $x_7$ -- $x_{10}$ are indicators for cases performed in 1990-1994, 1995--1999, 2000--2003 and 2004--2008 respectively.

\subsection{Model Fitting}

We fit the proposed GLMM model to the OPTN data, using a random effect following a Gaussian mixture distribution to represent the care quality of a center. 

Using the proposed test procedure to decide the order the latent Gaussian mixture model, the $p$-value is 0.0016 for $H_0: C_0=1$ vs. $H_1: C_0=2$; and 0.4076 for $H_0: C_0=2$ vs. $H_1: C_0=3$. We conclude that the care quality among the kidney transplant centers is not homogeneous and and the distribution of the random effect is adequately described by a two-component Gaussian mixture. The estimated fixed effects under our final model are summarized in Table \ref{tab:fixef}, where the standard errors are obtained using the asymptotic expansion (\ref{eq:asymp_bdt_n_c}).  As we can see, all covariates considered are significant. Since we code $Y=1$ as death, the results in Table \ref{tab:fixef} imply that patient death rate is higher if the donor kidney is not delivered to the patient fast enough, older patients have a higher death rate, men have higher death rate than women, and higher BMI also leads to higher risk. The coefficients for $x_7$ -- $x_{10}$ are negative and decreasing in their order confirming that the overall death rate is decreasing over time.

\begin{table}[ht]
\centering
\begin{tabular}{lrrrr}
  \hline 
 &      Estimate  &Std. Error& $z$-value &   $p$-value  \\
  \hline        
$x_1$ &   0.019503  & 0.0003048 & 63.9869&   $<$1e-99 \\
$x_2$ &  0.007112&  0.0002117  & 33.5890 &  $<$1e-99\\
$x_3$  &  0.030928 & 0.0094616 & 3.2688 &  0.0011\\
$x_4$  &  0.077860 &  0.0154998 & 5.0232 &   $<$1e-6\\
$x_5$  &   0.120536 &  0.0129628   &9.2986 &  $<$1e-19\\
$x_{6}$ &   0.225015  & 0.0148196 & 15.1836 &  $<$1e-51\\
$x_7$   & -0.270078 &  0.0146769 &-18.4016 &  $<$1e-74\\
$x_8$  &  -0.526297 &  0.0127432 &-41.3003 &  $<$1e-99\\
$x_9$  &  -0.632073 &  0.0138511 &-45.6334 &  $<$1e-99\\
$x_{10}$  &  -0.800276 &  0.0130163 &-61.4824 &  $<$1e-99\\

   \hline
\end{tabular}
\caption{OPTN data analysis: estimated GLMM regression coefficients, standard errors, $z$-values and $p$-values.  The covariates are $x_1$ =cold ischemic time, $x_2=$ age, $x_3=$ sex; $x_4$ -- $x_6$ are indicators for BMI in the intervals  (22, 25], (25-30] and 30+ respectively; $x_7$ -- $x_{10}$ are indicators for cases performed in 1990-1994, 1995--1999, 2000--2003 and 2004--2008 respectively. 
} 
\label{tab:fixef}
\end{table}

\begin{figure}[htb]
\centering \includegraphics[scale=0.5]{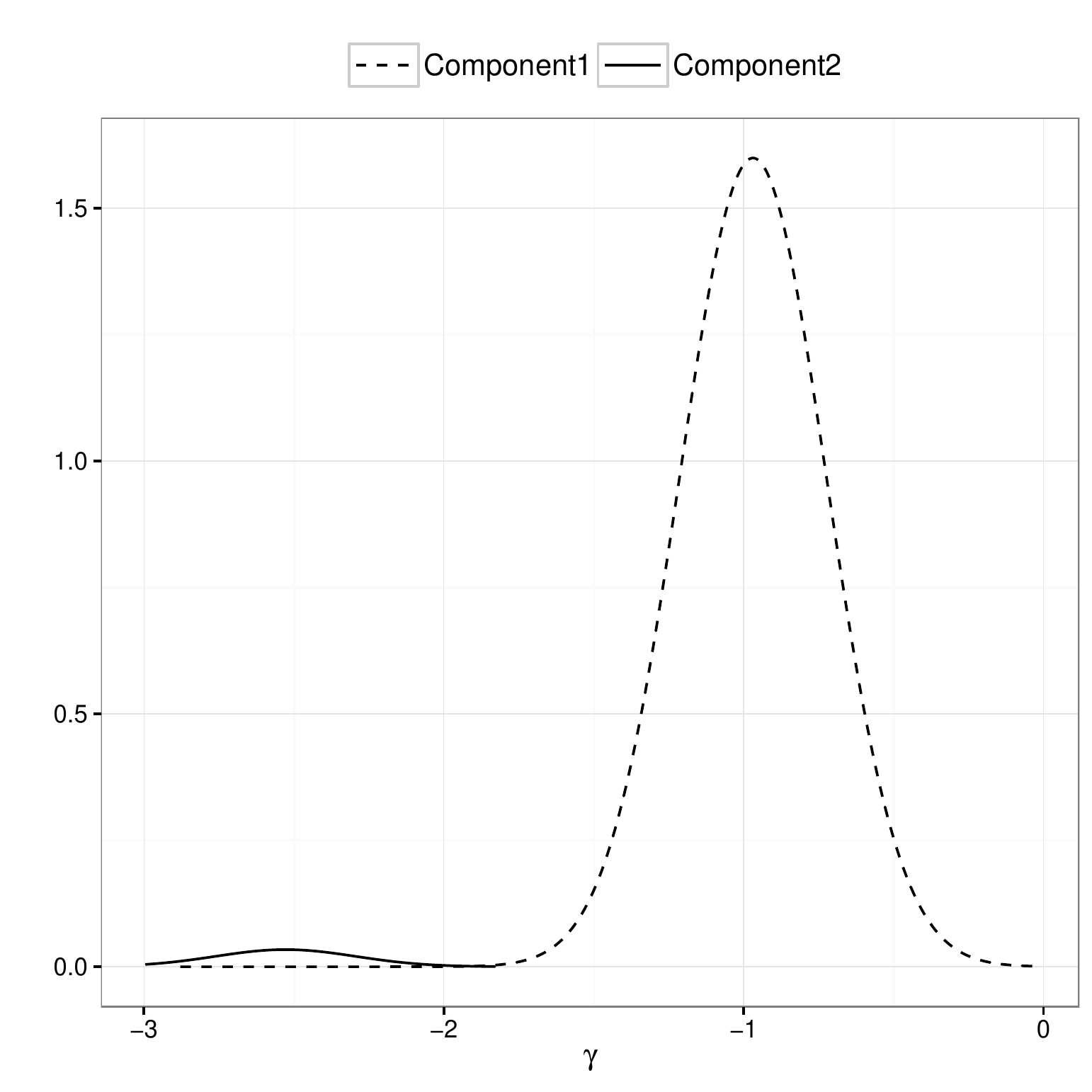}
\caption{Estimated latent Gaussian mixture model for the OPTN data. The ticks on the horizontal axis are the estimated random effects. \label{fig:density_realdata}}
\end{figure}

\begin{figure}[htb]
\centering \includegraphics[scale=0.5]{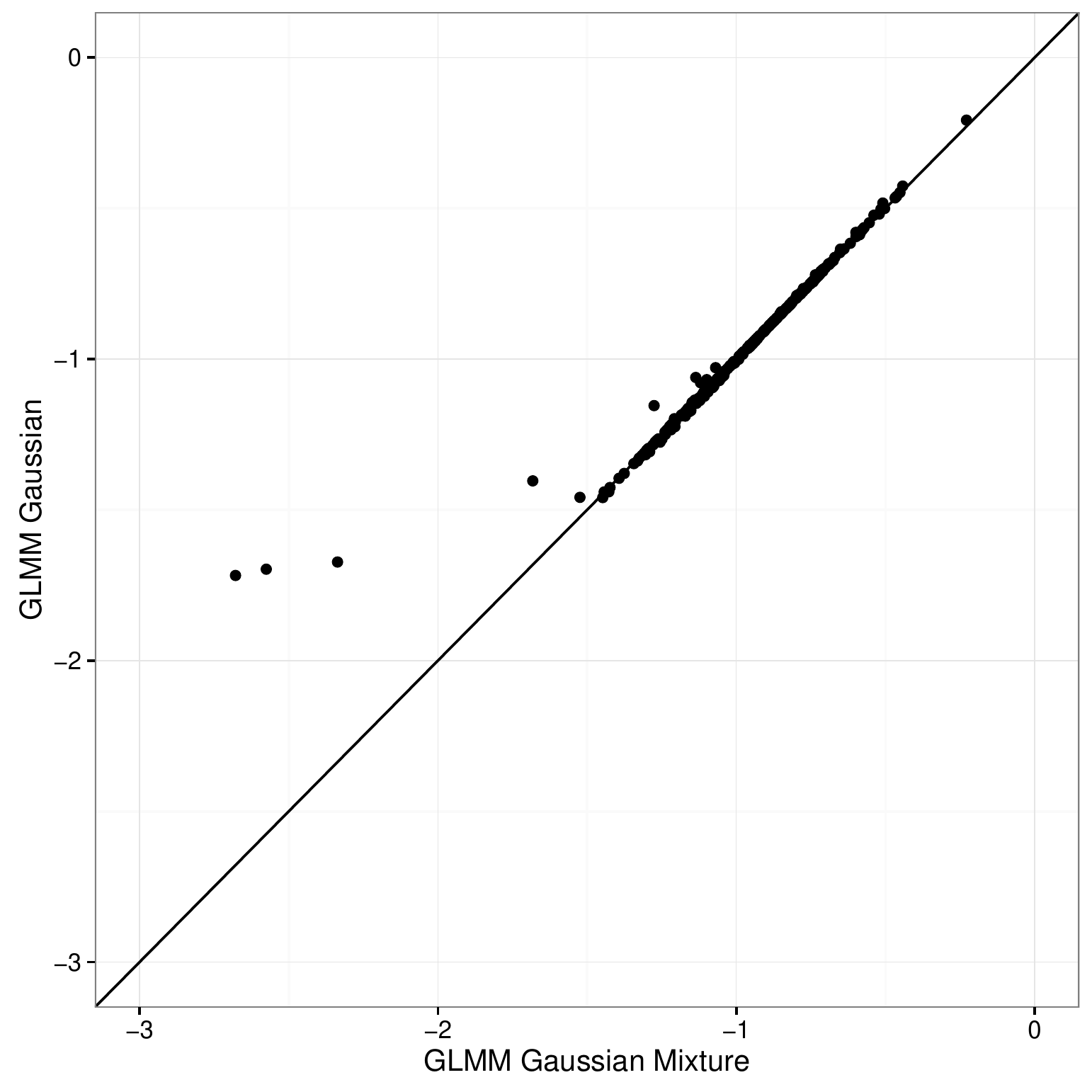}
\caption{Comparison of the predicted random effects in the OPTN data under Gaussian and Gaussian mixture model assumptions. \label{fig:real_gammacomparison12}}
\end{figure}

The estimated Gaussian mixture model for the random effect $\gamma$ is
$$ 0.98 \Normal(-0.969, 0.244^2) +
 0.02 \Normal(-2.528, 0.234^2). $$
The mixture density $g(\gamma)$ as well as the individual components are illustrated in Figure \ref{fig:density_realdata}. The majority of the centers have rather similar care quality, but there is also a small cluster of transplant centers that have lower death rate after taking into account of all the patient level covariates. These are the centers that are out-performing the others. In Figure \ref{fig:real_gammacomparison12}, we also compare the predicted random effects under GLMM with Gaussian random effects and those under our latent Gaussian mixture model. As we can see, for the majority of the centers, the predicted $\gamma$ is almost the same under both models, but, for the a few centers in the left tail, their care quality effects are seriously shrunk towards the mean if we assume the random effect follows a homogeneous Gaussian distribution.


Since the second component is small, we also run additional simulations to confirm that our methodology really works under such situations.
To mimic the real data, we simulate binary $Y_{ik}$ from a logistic GLMM using the covariates in the real data, set $\bdbeta$ at the estimated values in Table \ref{tab:fixef} and generate $\bdgamma$ from the following mixture model
$$ (1-\pi_2) \Normal(-0.969, 0.244^2) +
 \pi_2 \Normal(-2.528, 0.234^2).$$
We set $\pi_2$ to be 0.005, 0.01, 0.02 or 0.05, and simulate 200 data sets under each setting. The empirical powers for testing $H_0: C_0=1$ are 47\%, 78.5\%, 97.5\% and 100\% respectively. 
These results show that our method can detect a small component under the sample size of the real data and our discovery is likely to be true.

%


%
%
%
%
%
%
%
%
%
%
%
%
%
%
%
%
%
%

\subsection{Performance Evaluation}\label{sec:data_evaluation}
Based on the fitted model for $\gamma$ in Figure \ref{fig:density_realdata}, the majority of the centers provide similar care for their patients and the smaller component consists of transplant centers with lower mortality rate, which means these centers outperform the rest. We let the empirical null distribution to be the bigger component of the fitted mixture model. 
Using the evaluation procedure described in Section \ref{sec:fdr}, we find three transplant centers that outperforms the rest. In Table \ref{FDRids}, we list the id of the three outperforming centers, as well as their $lFDR$, $\wh \gamma$, number of cases treated, and their averaged 5-year survival rate.

\begin{table}[ht]
\centering
\begin{tabular}{rcccc}
  \hline     
 Center id & lFDR  &  $\hat{\gamma}$  &Sample Size &Survival Rate \\
      \hline

\#287 & 0.0013&      -2.6784&         114&        0.973\\
\#10  & 0.0061&      -2.5753&         125&         0.944\\
\#28  & 0.0736&      -2.3364&         120&        0.841\\

   \hline
\end{tabular}
\caption{The out-performing centers detected using local false discovery rate in the OPTN data.} 
\label{FDRids}
\end{table}

%
%
%
%
%
%
%
%
%
%
%
%
%
%
%
%
%
%

\section{SUMMARY}\label{sec:summary}
We propose a GLMM model with latent Gaussian mixture random effects that provides a natural framework to model the inhomogeneity among transplant centers and to rank their care quality. We demonstrate that the predicted random effects can be seriously shrunk toward the mean if the distribution of the random effect is mis-specified as Gaussian. This shrinkage effect is quite prominent for the centers in the tails of the population. The latent Gaussian mixture model is not strongly identifiable and suffers from slow convergence rate when the number of mixture component is larger than the truth. We develop test procedures to decide the number of mixture components.  Even though the proposed tests are designed mainly for testing scientific claims and providing uncertainty assessments, they can also be used for model selection and our simulation results in Section \ref{sec:simu_power} suggest the sequential test procedure outperforms a naive BIC. Developing a consistent model selection procedure for the latent Gaussian mixture model is our future work. The proposed test procedures are computationally intense, especially when analyzing large medical data sets like the OPTN data. This is because we have to try hundreds of initial values to find the biggest likelihood ratio. These computations are best handled using parallel computing. We have developed a software package \code{LatentGaussianMixtureModel} written in Julia (http://julialang.org/), which is a high-level, high-performance dynamic programming language. Our package is based on open source math libraries and supports parallel computing. We will make the package available on the correspondence author's website. Even though comparing transplant centers using five-year  survival rates of the patients has been the standard in the health policy literature, we acknowledge the fact that survival time is a more informative response variable. Extending the latent Gaussian mixture model to survival outcomes is also a topic for our future research.

%
%
%
%
%
%
%
%
%
%
%
%
%
%
%
%
%
%
\appendix
\renewcommand{\theequation}{A.\arabic{equation}}
\renewcommand{\thesubsection}{A.\arabic{subsection}}
\setcounter{equation}{0} \setcounter{figure}{0} \setcounter{lm}{0}

\section*{Appendix A: Simulation Approach for the Asymptotic Distribution in Proposition \ref{prop:TCdistn}}

We use the following procedure to simulate the asymptotic distribution in Proposition \ref{prop:TCdistn} under the hypothesis $H_0: C_0=C$.


\noindent{\bf Step 0.} Fit a $C$-component latent Gaussian mixture model and obtain the reduced model estimator $\wh{\bdtheta}_{red}$.

\noindent{\bf Step 1.} Calculate $\tilde{\bds}_i=(\bds_{\bdeta,i}\trans, \tilde{\bds}_{\boldsymbol{\lambda},i}\trans)\trans$ with $
	\tilde{\bds}_{\lambda, i} =
	\{(\bds_{\lambda, i}^{(1)} )\trans,
	(\bds_{\lambda, i}^{(2)} )\trans,
	\ldots 
	(\bds_{\lambda, i}^{(C)})\trans \}\trans
$, where $\bds_{\eta,i}$ and $\bds_{\lambda,i}^{(c)}$, $c=1,\ldots, C$, are the score functions for the restricted full models defined in (\ref{eq:score_C}) evaluated at $\wh{\bdtheta}_{red}$.
Let 
$$
	\tilde{\boldsymbol{I}}=\frac{1}{n}\sum_{i=1}^{n} {\tilde{\bds}}_{i}( {\tilde{\bds}}_{i})^{T}  
	=\left(\begin{array}{cc}
		\boldsymbol{I}_{\eta} & \tilde{\boldsymbol{I}}_{\eta\lambda}\\
		\tilde{\boldsymbol{I}}_{\lambda\eta} & \tilde{\boldsymbol{I}}_{\lambda}
	\end{array}\right).
$$
be the sample version of $\tilde{\boldsymbol{{\cal I}}}= E\tilde{\bds}_i\tilde{\bds}_i^{T}$, and calculate
$\tilde{\boldsymbol{I}}_{\lambda|\eta}=\tilde{\boldsymbol{I}}_{\lambda}-\tilde{\boldsymbol{I}}_{\lambda\eta}\boldsymbol{I}_{\eta}^{-1}(\tilde{\boldsymbol{I}}_{\lambda\eta})^{T}$.
To improve numerical stability, we check if $\tilde{\boldsymbol{I}}$ is an ill conditioned matrix. If so, set the eigenvalues with small absolute values to be a small positive number.

\noindent{\bf Step 2.}
Generate random a vector $\bds = \left\{ (\bds^{(1)})\trans,(\bds^{(2)})\trans, \ldots, (\bds^{(C)})\trans\right \} \trans \sim \Normal (0,\tilde{\boldsymbol{I}}_{\lambda|\eta})$. Let $\boldsymbol{I}_{\lambda|\eta}^{(c)}$ be the sub diagonal matrix of $\tilde{\boldsymbol{I}}_{\lambda|\eta}$ corresponding to $\bds^{(c)}$.
Then 
$$ T_C^\ast=\max\left\{(\bds^{(c)})^{T}(\boldsymbol{I}_{\lambda|\eta}^{(c)})^{-1}\bds^{(c)},c=1,2,\ldots,C\right\}$$
 has the same asymptotic distribution as $T_{C}(\tau)$ and $\wt T_C$.

\noindent{\bf Step 3.}
Repeat Step 2 a large number of times and use the empirical distribution of $T_C^\ast$ to approximate the asymptotic distribution of $\wt T_C$.

%
%
%
%
%
%
%
%
%
%
%
%
%
%
%
%
%
%
\baselineskip=14pt
\bibliographystyle{imsart-nameyear}
\bibliography{library,library2}

\clearpage\pagebreak\newpage
\baselineskip=18pt

\begin{center}
{\LARGE{\bf Supplementary Material to {\it Latent Gaussian Mixture Models for Nationwide Kidney Transplant Center Evaluation}}}
\end{center}

\vskip1cm 

\begin{center}
Lanfeng Pan, Yehua Li\\
Department of Statistics \& Statistical Laboratory, Iowa State University, Ames, IA 50011\\
\hskip 5mm \\
 Kevin He, Yanming Li and Yi Li \\
School of Public Health \& Kidney Epidemiology and Cost Center, University of Michigan, Ann Arbor, MI 48109. 
\par\end{center}

\vskip 5mm


\setcounter{equation}{0}
\setcounter{page}{1}
\setcounter{table}{1}
\setcounter{section}{0}
\renewcommand{\theequation}{S.\arabic{equation}}
\renewcommand{\thesection}{S.\arabic{section}}
\renewcommand{\thesubsection}{S.\arabic{section}.\arabic{subsection}}
\renewcommand{\thepage}{S.\arabic{page}}
\renewcommand{\thetable}{S.\arabic{table}}

\newtheorem{MyLemma}{ \underline{\smc Lemma}}[section]

\vskip1.5cm \noindent
The supplementary material contains the technical assumptions, proofs of the propositions and technical details of the EM algorithm.

\section{Assumptions and Consistency of the Estimator}\label{sec:assumption_lemma}
\setcounter{lm}{0}

\subsection{Assumptions}

For simplicity, assume $N_i=n_0$ for $i=1,\ldots, n$. 
Let $(\BX, \BY)$ be a generic copy of $(\BX_i, \BY_i)$ and have a density
\begin{equation}\label{eq:marg_density}
f(\bdx, \bdy | \bdtheta) = f(\bdx) \boldsymbol{\int} \left\{\prod_{k=1}^{n_0} f(y_{k}|\bdx_{k},\gamma;	\boldsymbol{\beta})
		g(\gamma|\bdtheta_{\gamma})\right\}d {\gamma}
\end{equation}
where $\bdy=(y_1,\ldots, y_{n_0})\trans$, $\bdx=(\bdx_1,\ldots, \bdx_{n_0})\trans$ and $f(\bdx)$ is the joint density
of $\BX$. Define metric 
$$
\delta(\bdtheta', \bdtheta)=\sum_l |\arctan \theta'_l -\arctan \theta_l|
$$
where $\theta_l$ is the $l$-th entry of $\bdtheta$. All convergences in the parameter space are defined with respect to $\delta$.

Assumptions 1- 5 below are equivalent to those in \citet{Kiefer1956} and \citet{Hathaway1985a} for the consistency result. Assumption 6 is a regularity assumption on the penalty function used in \cite{Chen2008a} and \cite{Kasahara2014}. Assumption 7 and 8 are additional assumptions for Propositions \ref{pp:convergencerate1} and \ref{pp:convergencerateC} respectively.

\begin{assumption}
$f(\bdx, \bdy | \bdtheta)$ is a density (the Radon-Nikodym derivative of a probability measure) with respect to a $\sigma$-finite measure $\mu$ on the space of $(\bdx,\bdy)$.
\end{assumption}

\begin{assumption}[Continuity Assumption] 
The definition of $f(\bdx, \bdy | \bdtheta)$ can be extended to the closure of the parameter space $\bar{\Theta}_C$ such that, for any $\bdtheta^*$ in $\bar{\Theta}_C$ and any Cauchy sequence $\{\bdtheta_1,\bdtheta_2,\ldots\}\subset \bar\Theta_C$, 
$
f(\bdx,\bdy | \bdtheta_i) \rightarrow f(\bdx, \bdy | \bdtheta^*)
$
if $\bdtheta_i \rightarrow \bdtheta^*$.
\end{assumption}

\begin{assumption}
For any $\bdtheta \in \bar{\Theta}_C$ and any $\rho > 0$, $\omega(\bdx, \bdy|\bdtheta, \rho)$ is a measurable function of $(\bdx, \bdy)$, where 
$$
\omega(\bdx, \bdy|\bdtheta, \rho) = \sup f(\bdx, \bdy | \bdtheta'),
$$
the supreme being taken over all $\bdtheta'$ in $\bar{\Theta}_C$ for which $\delta(\bdtheta', \bdtheta)<\rho$.
\end{assumption}

\begin{assumption}[Identifiability Assumption] 
Identify $\bar\Theta_C$ as the quotient topological space such that $\CF$ defined in (\ref{eq:equiv_param}) is identified as a single point.
\end{assumption}

\begin{assumption}
For any $\bdtheta'$ in $\bar{\Theta}_C$,
$$
\lim_{\rho \downarrow 0} E_\bdtheta \left[ \log \frac{\omega(\bdx, \bdy|\bdtheta', \rho)}{f(\bdx, \bdy|  \bdtheta)} \right]^+ < \infty,
$$
where $E_\bdtheta$ is the expectation under $f(\bdx,\bdy|\bdtheta)$.
\end{assumption}


\begin{assumption}
The penalty function satisfies, (a) $\sup_{\sigma^2>0} \max\{ 0, p_n(\sigma^2)\}=o(n)$, $p_n(\sigma^2)=o(n)$ for any fixed $\sigma^2$; (b) for any $\sigma \in (0, 8/(nM)]$, $p_n(\sigma^2) \le 5 \{ln(n)\}^2 \ln (\sigma)$ for sufficient large $n$, where $M=\sup_{\bdx,\bdy} f(\bdy|\bdx; \bdtheta_0)$; (c) $p_n'(\sigma^2)=o_p(n^{1/4})$ for any fixed $\sigma^2$. 
\end{assumption}

\begin{assumption}
When the true number of component is $C_0=1$, assume that $\boldsymbol{{\cal I}}=E\boldsymbol{I}_{n}$ is a finite, positive definite matrix, where $\boldsymbol{I}_{n}$ is defined in (\ref{eq:local_quadratic_like}).
\end{assumption}

\begin{assumption}
When $\bdtheta\in \Theta_C$, assume that $\boldsymbol{{\cal I}}^{(c)}$ defined in (\ref{eq:BCI_c}) is positive definite, for $c=1,2,\ldots, C$.
\end{assumption}

\noindent{\bf Remarks:}

\noindent{\bf 1.} The continuity assumption (Assumption 2) is not satisfied by the finite Gaussian mixture model on the boundary of the parameters space, since the likelihood diverges $\infty$ if any $\sigma_c^2\to 0$. That is the reason that Hathaway (1985) restricted the estimation in the interior of the parameter space. However, in our problem, the finite Gaussian mixture density $g(\gamma)$ is convoluted with proper density $f(\bdy|\bdx,\gamma)$ in (\ref{eq:marg_density}). Since the integral is bounded, unbounded likelihood is no longer a concern and the condition is satisfied even on boundary points of $\bar\Theta_C$. 

\noindent{\bf 2.} Assumption 4 is a modified version of the identifiability assumption in Kiefer and Wolfowitz (1956). The same assumption is used in Hathaway (1985). The consistency result in Proposition \ref{prop:consistency} means consistently estimating the mixture density.

\subsection{Proof of Proposition \ref{prop:consistency}}

Using similar arguments as in \citet{Chen2008a} one can show, as long as the penalty function satisfies Assumption 6, the maximizer of (\ref{eq:pen_marg_like}) is restricted in an interior region of the parameter space $\bar\Theta(\epsilon)=\{\bdtheta \in \bar\Theta; \min_c \sigma_c^2\ge \epsilon\}$ for some positive constant $\epsilon$. Since the penalty term is of order $o(n)$, which is much smaller than the likelihood function, the maximum penalized likelihood estimator $\wh\bdtheta$ in the restricted parameter space belong to the class of {\it modified maximum likelihood estimator} in \citet{Kiefer1956} and the strong consistency of $\wh\bdtheta$ follows from their theory.

\section{Proof of Proposition \ref{pp:convergencerate1}}\label{sec:proof_conv_rate1}

Denote for convenience
$
\zeta_{i} = \prod_{k=1}^{n_0} f(y_{ik}|\bdx_{ik},\gamma_{i};	\bdtheta_y).
$
After fixing $\pi_1=\tau$, the log likelihood is 
\begin{eqnarray*}
l_n(\bdtheta) 
&=&\sum_{i=1}^n\log\int\zeta_{i}\{\tau f_1(\gamma|\mu_{1},\sigma_{1})+(1-\tau)f_2(\gamma|\mu_{2},\sigma_{2})\}d\gamma.
\end{eqnarray*}

We adopt the re-parameterization of \citet{Kasahara2014}, 
\ben\label{eq:reparam1}
	\left(\begin{array}{c}
	\mu_{1}\\
	\mu_{2}\\
	\sigma_{1}^{2}\\
	\sigma_{2}^{2}
	\end{array}\right)=\left(\begin{array}{c}
	\nu_{\mu}+(1-\tau)\lambda_{\mu}\\
	\nu_{\mu}-\tau\lambda_{\mu}\\
	\nu_{\sigma}+(1-\tau)(2\lambda_{\sigma}-\frac{1+\tau}{3}\lambda_{\mu}^{2})\\
	\nu_{\sigma}-\tau(2\lambda_{\sigma}+\frac{2-\tau}{3}\lambda_{\mu}^{2})
	\end{array}\right),
\een
collect all parameters except $\tau$ into $\bdpsi(\tau)=(\bdeta\trans,\boldsymbol{\lambda}\trans)\trans$, where $\bdeta=(\bdtheta_y\trans,\nu_{\mu},\nu_{\sigma})\trans$
and $\boldsymbol{\lambda}=(\lambda_{\mu},\lambda_{\sigma})\trans$. Denote $\bar{\Theta}_{\psi}(\tau)$ as the parameter space of $\bdpsi$ corresponding to $\bar{\Theta}_2(\tau)$. Sometimes we suppress the dependence of $\bdpsi(\tau)$ on $\tau$. Under the null hypothesis $C_0=1$, $\lambda_\mu=\lambda_\sigma=0$ and the true parameter vector is $\bdpsi^{*}=((\bdeta^{*})\trans,0,0)\trans$. 

For any multivariate function $f(\bdx)$, denote $\nabla_{\bdx^k} f$ as its $k$-th derivative, which is a multidimensional array. By similar calculations as in Proposition C and equation (29) in the supplementary appendix of \citet{Kasahara2014}, we can show
\bse
	&& \nabla_{\lambda_{\mu}^k, \bdeta^\ell} l_n(\bdpsi^\ast,\tau)=0, \quad \hbox{ for $k=1,2,3$ and $ \ell=0,1,2\ldots$};\\
	&&\nabla_{\lambda_\mu^k} l_n(\bdpsi^\ast,\tau)=O_p(n^{1/2}), \quad \hbox{ for $k=4,5,6,7$};\\
	&&\nabla_{\lambda_{\sigma} \bdeta^\ell,\tau} l_n(\bdpsi^\ast)=0, \quad \hbox{ for $\ell=0,1,2,\ldots$};\\
	&&\nabla_{\lambda_\sigma^k} l_n(\bdpsi^\ast,\tau)=O_p(n^{1/2}), \quad \hbox{ for $k=2,3$};\\
	&&\nabla_{\lambda_\mu \lambda_\sigma^2} l_n(\bdpsi^\ast,\tau)=O_p(n^{1/2});\\
	&&\nabla_{\lambda_\mu^k \lambda_\sigma} l_n(\bdpsi^\ast,\tau)=O_p(n^{1/2}), \quad \hbox{ for $k=1,\ldots,4$}.
\ese

Denote $g^\ast(\gamma)=g(\gamma; \bdpsi^\ast)$ as the true density of $\gamma$ under the null hypothesis. Using a ninth order Taylor expansion of $l_{pen}$ around $\bdpsi^\ast$ as in \citet{Kasahara2014}, we get
the following local quadratic approximation to the penalized likelihood
\ben\label{eq:local_quadratic_like}
	l_{pen}(\bdpsi,\tau)-l_{pen} (\bdpsi^{*}, \tau) &=& \bdt_{n}(\bdpsi,\tau)\trans\boldsymbol{S}_{n}-\frac{1}{2}\bdt_{n}(\bdpsi,\tau)\trans\boldsymbol{I}_{n}\bdt_{n}(\bdpsi,\tau)+R_{n}(\bdpsi,\tau) \nonumber \\
	&& \hskip20mm + \sum_{c=1}^2 [p_n\{ \sigma_c^2(\bdpsi, \tau)\}- p_n\{\sigma_c^2(\bdpsi^\ast,\tau)\}] ,
\een
where $\bdt_{n}(\bdpsi,\tau)=(\bdt_{\bdeta,n},\bdt_{\boldsymbol{\lambda},n})\trans$, $\boldsymbol{S}_{n}=\frac{1}{\sqrt{n}}\sum_{i=1}^{n}\boldsymbol{s_{i}}$, $\boldsymbol{I}_{n}=\frac{1}{n}\sum_{i=1}^{n}\boldsymbol{s_{i}}\boldsymbol{s_{i}}\trans$, $\bds_{i}=(\bds_{\bdeta,i}\trans,\bds_{\boldsymbol{\lambda},i}\trans)\trans$, $\sigma_c^2(\bdpsi, \tau)$ is the variance as a function of $\bdpsi$ defined by the reparameterization in (\ref{eq:reparam1}),
\bse
	&&\bdt_{\bdeta,n}=\sqrt{n}(\bdeta-\bdeta*), \quad  
	\bdt_{\boldsymbol{\lambda},n}=\left(\begin{array}{c}
	6\sqrt{n}\tau(1-\tau)\lambda_{\mu}\lambda_{\sigma}\\
	\sqrt{n}\tau(1-\tau)(12\lambda_{\sigma}^{2}-\frac{2}{3}(\tau^{2}-\tau+1)\lambda_{\mu}^{4})
	\end{array}\right),\\
	&&\bds_{\bdeta,i}=\left(\begin{array}{c}
	\bds_{\boldsymbol{\theta_y},i}\\
	s_{\nu_{\mu},i}\\
	s_{\nu_{\sigma}, i}
	\end{array}\right)=\left(\begin{array}{c}
	\frac{\int (\partial \zeta_{i}/ \partial \bdtheta_y) g^{*}}{\int\zeta_{i}g^{*}}\\
	\frac{\int\zeta_{i}g^{*}H_{i}^{1*}}{\int\zeta_{i}g^{*}}\\
	\frac{\int\zeta_{i}g^{*}H_{i}^{2*}}{\int\zeta_{i}g^{*}}
	\end{array}\right), \quad 
	\bds_{\boldsymbol{\lambda},i}=\left(\begin{array}{c}
	\frac{\int\zeta_{i}g^{*}H_{i}^{3*}}{\int\zeta_{i}g^{*}}\\
	\frac{\int\zeta_{i}g^{*}H_{i}^{4*}}{\int\zeta_{i}g^{*}}
	\end{array}\right), \\
	&& R_{n}(\bdpsi,\tau)=[O(\| \bdpsi - \bdpsi^* \|) +o(1) ] \times O_p[\{1+\| \bdt_n(\bdpsi, \tau) \|^2 \}].
\ese
Here,
\[
	H_{i}^{k*}=H^{k}(\frac{\gamma_{i}-\mu_\gamma^*}{\sigma_\gamma^*})/(k!(\sigma_\gamma^*)^{k})
\]
where $H^k(x)$ is the $k$th order Hermite polynomial, e.g. $H^0(x)=1$, $H^1(x)=x$, $H^2(x)=x^2-1$, $H^3(x)=x^3-3x$ and $H^4(x)=x^4-6x^2+3$.

By consistency of the estimator, we can focus on $\bdpsi$ such that $\|\bdpsi-\bdpsi^\ast\|=o_p(1)$ and hence $R_n(\bdpsi,\tau)=o_p(\|\bdt_n(\bdpsi, \tau)\|^2)$. By Assumption 6, $p_n'(\sigma^2)=o_p(n^{1/4})$, and by (\ref{eq:reparam1})
\begin{eqnarray*}
 p_n\{ \sigma_c^2(\bdpsi, \tau)\}- p_n\{\sigma_c^2(\bdpsi^\ast,\tau)\} = o_p(n^{1/4})(|\lambda_\sigma| +\lambda_\mu^2) = o_p\{\|\bdt_n(\bdpsi,\tau)\|\}.
\end{eqnarray*}
Therefore, $l_{pen}(\bdpsi,\tau)-l_{pen} (\bdpsi^{*}, \tau)$ is dominated by the quadratic function defined by the first two terms on the right hand side of (\ref{eq:local_quadratic_like}). It is then easy to see $\wh\bdt_n= \bdt_n\{\wh\bdpsi(\tau),\tau\}$ that maximizes $l_{pen}(\bdpsi,\tau)-l_{pen} (\bdpsi^{*}, \tau)$ is given by 
\ben\label{eq:asymp_t_n}
	\wh \bdt_n=  \BI_n^{-1} \BS_n +o_p(1).
\een
Under Assumption 7, $\boldsymbol{{\cal I}}=E\boldsymbol{I}_{n}$ is a positive definite matrix. By the law of large numbers, $\BI_n\to \BCI$ in probability. On the other hand, by the central limit theorem, $\BS_n \to \Normal (\pmb{0}, \BCI)$ in distribution. Therefore, $\wh \bdt_n\to \Normal(\pmb{0}, \BCI^{-1})$ in distribution, which also implies
\bse
	\wh \bdbeta_{full}(\tau)-\bdbeta_0= O_p(n^{-1/2}), \quad \wh\lambda_\mu=O_p(n^{-1/4}), \quad \hbox{and   } \wh \lambda_\sigma=O_p(n^{-1/4}).
\ese
The convergence rate of $\wh\bdtheta_{\gamma,full}(\tau)$ is determined by those of $\wh\lambda_\mu$ and $\wh\lambda_\sigma$.

\section{Proof of Proposition \ref{prop:T1distn}}\label{sec:proof_T1}
Following arguments in Section \ref{sec:proof_conv_rate1}, we have $$\BS_n \to \Normal (\pmb{0}, \BCI)$$ in distribution, where  $\boldsymbol{{\cal I}}=E\boldsymbol{I}_{n}$. Under the full model, for any $\bdpsi$ such that $\bdt_n=O_p(1)$, using the local quadratic approximation (\ref{eq:local_quadratic_like}) we have
\begin{eqnarray*}
	2\{l_n(\bdpsi,\tau)-l_n(\bdpsi^{*},\tau)\} &=& 2\bdt_{n}\trans\boldsymbol{S}_{n}-\bdt_{n}\trans\boldsymbol{I}_{n}\bdt_{n}+o_p(1) \\
	&=& 2 \bdt_{n}\trans\boldsymbol{S}_{n}-\bdt_{n}\trans\boldsymbol{\mathcal{I}}\bdt_{n}+o_{p}(1).
\end{eqnarray*}
Let $\wh \bdpsi_{full}(\tau)$ be maximizer of (\ref{eq:local_quadratic_like}) under the full model with 2 components, and it is the reparameterized version of $\wh\bdtheta_{full}(\tau)$. By (\ref{eq:asymp_t_n}), $\bdt_n\{\wh\bdpsi_{full}(\tau)\} =\BCI^{-1} \BS_n+o_p(1)$ and hence
\ben\label{eq:max_like_full}
	2[l_n\{\wh\bdpsi_{full}(\tau),\tau\}-l_n(\bdpsi^{*},\tau)]= \BS_n\trans \BCI^{-1} \BS_n +o_p(1).
\een

Partition $\BS_n$ into 
$\left(\begin{array}{c}
\BS_{\eta, n}\\
\BS_{\lambda, n}
\end{array}\right)$ 
according to the partition of $\bdpsi$. With a similar partition to $\BCI$, we have
\bse
	\BCI^{-1}=
	\left(\begin{array}{cc}
	\boldsymbol{\mathcal{I}}_{\eta} & \boldsymbol{\mathcal{I}}_{\eta\lambda}\\
	\boldsymbol{\mathcal{I}}_{\lambda\eta} & \boldsymbol{\mathcal{I}}_{\lambda}
	\end{array}\right)^{-1}
	=\left(\begin{array}{cc}
	\BCI_\eta^{-1} +\BCI_\eta^{-1} \BCI_{\eta \lambda} \BCI_{\lambda|\eta}^{-1} \BCI_{ \lambda\eta } \BCI_\eta^{-1}
	& -\boldsymbol{\mathcal{I}}_{\eta}^{-1}\boldsymbol{\mathcal{I}}_{\eta\lambda}\boldsymbol{\mathcal{I}}_{\lambda|\eta}^{-1}\\
	(-\boldsymbol{\mathcal{I}}_{\eta}^{-1}\boldsymbol{\mathcal{I}}_{\eta\lambda}\boldsymbol{\mathcal{I}}_{\lambda|\eta}^{-1})\trans & \boldsymbol{\mathcal{I}}_{\lambda|\eta}^{-1}
	\end{array}\right),
\ese
where $\boldsymbol{\mathcal{I}}_{\lambda|\eta}=\boldsymbol{\mathcal{I}}_{\lambda}-\boldsymbol{\mathcal{I}}_{\lambda\eta}\boldsymbol{\mathcal{I}}_{\eta}^{-1}\boldsymbol{\mathcal{I}}_{\eta\lambda}$. 
Define
\bse
	\BS_{\lambda|\eta, n}= \BS_{\lambda, n}- \BCI_{\lambda \eta} \BCI_{\eta}^{-1} \BS_{\eta, n},
\ese
and by simple algebra
\ben\label{eq:quadratic_S_n}
	\BS_n\trans \BCI^{-1} \BS_n= \BS_{\eta, n}\trans \BCI_\eta^{-1} \BS_{\eta, n} +\BS_{\lambda|\eta, n}\trans \BCI_{\lambda|\eta}^{-1} \BS_{\lambda | \eta, n}.
\een

Under the reduced model, $\bdlambda={\pmb 0}$, and hence $\bdt_{\lambda n}=\BS_{\lambda n}={\pmb 0}$. Using the same local quadratic approximation, for a parameter vector $\bdpsi_{red}$ in the reduced model,
\begin{eqnarray*}
	2\{l_n(\bdpsi_{red},\tau)-l_n(\bdpsi^{*},\tau)\} = 2\bdt_{\eta n}\trans\boldsymbol{S}_{\eta n}-\bdt_{\eta n}\trans\boldsymbol{I}_{\eta} \bdt_{\eta n}+o_p(1).
\end{eqnarray*}
Let $\wh\bdpsi_{red}$ be the estimator that maximizes the reduced model penalized likelihood, then $\bdt_{\eta n} (\wh\bdpsi_{red})= \BCI_{\eta}^{-1} \BS_{\eta n} +o_p(1)$, and 
\ben\label{eq:max_like_red}
	2\{ l_n(\wh\bdpsi_{red},\tau)-l_n(\bdpsi^{*},\tau)\} = \BS_{\eta, n}\trans \BCI_\eta^{-1} \BS_{\eta, n} +o_p(1).
\een
Combining (\ref{eq:max_like_full}), (\ref{eq:quadratic_S_n}) and (\ref{eq:max_like_red}),
\bse
	T_1(\tau)=2[l_n\{\wh\bdpsi_{full}(\tau),\tau\}-l_n(\wh\bdpsi_{red},\tau)] = \BS_{\lambda|\eta, n}\trans \BCI_{\lambda|\eta}^{-1} \BS_{\lambda | \eta, n} +o_p(1) \cid \chi^2(2).
\ese

Because $\BS_{\lambda|\eta, n}$ and $ \BCI_{\lambda|\eta}$ do not depend on $\tau$,
\bse
	\wt T_1=\max_{\tau\in \BCT} T_1(\tau)= \BS_{\lambda|\eta, n}\trans \BCI_{\lambda|\eta}^{-1} \BS_{\lambda | \eta, n} +o_p(1) \cid \chi^2(2).
\ese

\section{Proof of Proposition \ref{pp:convergencerateC}}

Denote 
$
\zeta_{i} = \prod_{k=1}^{n_0} f(y_{ik}|\bdx_{ik},\gamma_{i};	\bdtheta_y)
$
as in Section \ref{sec:proof_conv_rate1}. Under the local reparameterization in $\CN_{C+1}(c,\tau)$ defined in (\ref{eq:reparam_C}) and (\ref{eq:reparam_C_rest}) in Section \ref{sec:high_order_test}, the log likelihood is
\begin{eqnarray*}
	l_n(\bdtheta) &=&\sum_{i=1}^n\log\int\zeta_{i}g_{c,\tau}(\gamma)d\gamma
\end{eqnarray*}
where
\begin{eqnarray*}
	g_{c,\tau}(\gamma) &=& (\pi_{c}+\pi_{c+1})\tau f(\gamma |\mu_{c},\sigma_{c})+ (\pi_{c}+\pi_{c+1})(1-\tau)f(\gamma|\mu_{c+1},\sigma_{c+1}) \\
	&& +\sum_{c'\neq c}\pi_{c'}f_{c'}(\gamma\mid \mu_{c'}, \sigma_{c'}) \\
	&=&
	(\pi_{c}+\pi_{c+1})\tau f\left\{\gamma|\nu_{\mu}+(1-\tau)\lambda_{\mu},\nu_{\sigma}+(1-\tau)(2\lambda_{\sigma}-\frac{1+\tau}{3}\lambda_{\mu}^{2})\right\}
	\\
	&& +(\pi_{c}+\pi_{c+1})(1-\tau)f\left\{\gamma|\nu_{\mu}-\tau \lambda_{\mu},\nu_{\sigma}
	-\tau (2\lambda_{\sigma}+\frac{2-\tau}{3}\lambda_{\mu}^{2})\right\} \\
	&& +\sum_{c' \neq c}\pi_{c'}f_{c'}(\gamma\mid \mu_{c'}, \sigma_{c'}).
\end{eqnarray*}

The score function with respect to $\bdpsi(c, \tau)$ is
$
\bds^{(c)}_i = (\bds_{\bdeta,i}\trans, (\bds_{\boldsymbol{\lambda},i}^{(c)})\trans)\trans$, which is defined in (\ref{eq:score_C}).
Define $\boldsymbol{S}_n^{(c)}=\frac{1}{\sqrt{n}}\sum_{i=1}^{n}\bds_{i}^{(c)}$, $\boldsymbol{I}_n^{(c)}=\frac{1}{n}\sum_{i=1}^{n}\bds_{i}^{(c)}(\bds_{i}^{(c)})\trans$ and
$\bdt_{n}(\bdpsi(c,\tau),\tau)=(\bdt_{\bdeta,n},\bdt_{\boldsymbol{\lambda},n})\trans$ where
\bse
	&&\bdt_{\bdeta,n}=\sqrt{n}(\bdeta-\bdeta*), \quad  
	\bdt_{\boldsymbol{\lambda},n}=\left(\begin{array}{c}
	6\sqrt{n}\tau(1-\tau)\lambda_{\mu}\lambda_{\sigma}\\
	\sqrt{n}\tau(1-\tau)(12\lambda_{\sigma}^{2}-\frac{2}{3}(\tau^{2}-\tau+1)\lambda_{\mu}^{4})
	\end{array}\right).
	%
	%
	%
	%
	%
%
\ese
Similar to (\ref{eq:local_quadratic_like}), we can derive a local quadratic approximation to the likelihood 
\begin{eqnarray}\label{eq:local_quadratic_like_C}
	l_n(\bdpsi(c, \tau),\tau)-l_n(\bdpsi^*) &=& \bdt_n(\bdpsi(c,\tau), \tau)\trans\boldsymbol{S}^{(c)}_{n} 
	-\frac{1}{2}\bdt_n(\bdpsi(c,\tau), \tau)\trans\boldsymbol{I}_{n}^{(c)}\bdt_n(\bdpsi(c,\tau), \tau) \nonumber \\
	&&+R_{n,c}(\bdpsi(c, \tau), \tau).
\end{eqnarray}
where $ R_{n}(\bdpsi,\tau)=[O(\| \bdpsi - \bdpsi^* \|) +o(1) ] \times O_p[\{1+\| \bdt_n(\bdpsi, \tau) \|^2 \}]$.

Put $\wh{\bdpsi}_{full} (c,\tau)=\arg\max_{\bdpsi(c, \tau)\in \Theta_{\psi}(c, \tau)} l_{pen} \left(\bdpsi(c, \tau),\tau\right)$ and $\wh{\bdt}_n=\bdt_n\left(\wh{\bdpsi}_{full}(c,\tau),\tau\right)$. Using similar arguments as in Section \ref{sec:proof_conv_rate1}, we can show that the penalty function is asymptotically negligible when $\bdpsi(c, \tau)$ is in a consistent neighborhood of $\bdpsi^\ast$.
Define 
\ben\label{eq:BCI_c}
	\BCI^{(c)}=E( \BI_n^{(c)}) = \var(\bds_i^{(c)}), 
\een
which is positive definite under Assumption 8. It is then easy to see that 
\ben\label{eq:asymp_bdt_n_c}
	\wh \bdt_n= (\BCI^{(c)} )^{-1} \BS_n^{(c)} +o_p(1) \cid \Normal \{0, (\BCI^{(c)} )^{-1} \}.
\een
By the definition of $\bdt_n\{\bdpsi(c,\tau),\tau\}$, we get $\wh\bdeta-\bdeta^\ast=O_p(n^{-1/2})$, $\wh\lambda_\mu=O_p(n^{-1/4})$ and $\wh \lambda_\sigma=O_p(n^{-1/4})$. Since the convergence rates for $\wh\mu_{c,full}(c,\tau)$, $\wh\mu_{c+1,full}(c,\tau)$, $\wh\sigma_{c,full}(c,\tau)$ and $\wh\sigma_{c+1,full}(c,\tau)$ are determined by $\wh\lambda_\mu$ and $\wh\lambda_\sigma$, they converge to the true parameters in a slower $O_p(n^{-1/4})$ rate and the rest of the parameters in $\wh\bdtheta_{full}(c,\tau)$ converge in a $O_p(n^{-1/2})$ rate.



%
%
%
%
%
%
%
%
%
%
%

\section{Proof of Proposition \ref{prop:TCdistn}}\label{sec:proof_TC}
We first derive the asymptotic properties for $T_C(c,\tau)$. By (\ref{eq:local_quadratic_like_C}) and (\ref{eq:asymp_bdt_n_c}),
\bse
	2[ l_n\{\wh{\bdpsi}_{full}(c, \tau),\tau\}-l_n(\bdpsi^*)] = (\BS_n^{(c)})\trans (\BCI^{(c)}) ^{-1} \BS_n^{(c)} +o_p(1),
\ese
where $\BS_n^{(c)}\cid \Normal({\pmb 0}, \BCI^{(c)})$ by the central limit theorem.

Note that the reduced model estimator $\wh\bdpsi_{red}(c,\tau)$ is obtained by minimizing the penalized likelihood while restricting $\lambda_\mu=\lambda_\sigma=0$. by similar derivations under the full model, we get
\bse
	2[ l_n\{\wh{\bdpsi}_{red}(c, \tau),\tau\}-l_n(\bdpsi^*)] = \BS_{\eta, n}\trans \BCI_\eta^{-1} \BS_{\eta, n} +o_p(1),
\ese
where $\BS_{\eta, n}$ and $\BCI_\eta$ are sub-vector or sub-matrix of $\BS_n^{(c)}$ and $\BCI^{(c)}$ as defined in Proposition \ref{prop:TCdistn}.

Using algebra similar to that in Section \ref{sec:proof_T1}, we get 
\begin{eqnarray*}
	T_C(c,\tau)&=&2[ l_n\{\wh{\bdpsi}_{full}(c, \tau),\tau\}-l_n\{\wh\bdpsi_{red}(c,\tau),\tau\}] \\
	&=& (\boldsymbol{S}_{\lambda|\eta, n}^{(c)})\trans(\boldsymbol{{\cal I}}_{\lambda|\eta}^{(c)})^{-1}\boldsymbol{S}_{\lambda|\eta, n}^{(c)}+o_p(1) \\
	&\cid& \chi^2(2).
\end{eqnarray*}
Therefore,
\bse
	T_C(\tau)=\max_c T_C(c,\tau)\cid \max\{ (\boldsymbol{S}_{\lambda|\eta, n}^{(c)})\trans(\boldsymbol{{\cal I}}_{\lambda|\eta}^{(c)})^{-1}\boldsymbol{S}_{\lambda|\eta, n}^{(c)}, \quad c=1,\ldots,C\}.
\ese
Since none of the quantities $(\boldsymbol{S}_{\lambda|\eta, n}^{(c)})\trans(\boldsymbol{{\cal I}}_{\lambda|\eta}^{(c)})^{-1}\boldsymbol{S}_{\lambda|\eta, n}^{(c)}$ depends on $\tau$, $\wt T_C$ that maximizes $T_C(\tau)$ over any set $\CT$ has the same limiting distribution.

\section{Proof of Proposition \ref{prop:FDR}}\label{sec:proof_fdr}

The FDR for the described procedure is
\bse
FDR&=&E\left\{\frac{\sum_i^n I(\delta_i=1,\sum_{c\in \CC_0}L_{ic}=1) }{\sum_i^n I(\delta_i=1)}  \Big| \sum_i^n I(\delta_i=1) > 0\right\} P\left\{\sum_i^n I(\delta_i=1) > 0\right\} \\
&=& E\left\{\frac{\sum_i^n \delta_i \left(\sum_{c\in \CC_0}L_{ic}\right) }{\sum_i^n \delta_i \vee 1} \right\} \\
&=& E\left\{\frac{\sum_i^n \delta_i E\left(\sum_{c\in \CC_0}L_{ic} =1 \big| \BX_i,\BY_i \right)}{\sum_i^n \delta_i \vee 1}  \right\} \\
&=& E\left(\frac{\sum_i^n \delta_i lFDR_i }{\sum_i^n \delta_i \vee 1}  \right) \\
&=& E \left(\frac{\sum_i^k lFDR_{(i)} }{k}  \right) \\
&\le & \alpha.
\ese

\section{Computation Details}\label{sec:computation}

We now provide more details on the {Gauss-Hermite Approximation} used in Section \ref{estimation}. The EM loss function is
\begin{eqnarray}
	Q(\bdtheta|\bdtheta^{(t-1)}) 
	= \sum_{i=1}^n E\left[\ell_{i, comp}(\theta; \BY_i , \BX_i, \gamma_i, \BL_i)|\BY_i,\BX_i,\bdtheta^{(t-1)}\right]  +\sum_{c=1}^Cp_n(\sigma^{2}_{c};\wh\sigma^2_{pilot}) 
\end{eqnarray}
where 
\bse
   	&& E\left[\ell_{i, comp}(\theta; \BY_i , \BX_i, \gamma_i, \BL_i)|\BY_i,\BX_i,\bdtheta^{(t-1)}\right]   \\
 	&& \hskip2cm = \sum_{c=1}^C \int \log f(\BY_{i}|\BX_{i},\gamma;\bdtheta_y) f(\gamma, L_{ic}=1|\BX_{i}, \BY_{i}; \bdtheta^{(t-1)}) d\gamma   \\
	&& \hskip2.5cm +\sum_{c=1}^C\int\left[\log 
	\{\phi\left(\frac{\gamma-\mu_c}{\sigma_c}\right)/\sigma_c\}f(\gamma, L_{ic}=1|\BX_{i}, \BY_{i}; \bdtheta^{(t-1)})\right]d\gamma  \\
  	&& \hskip2.5cm+\sum_{c=1}^C\log\pi_c\int f(\gamma, L_{ic}=1|\BX_{i}, \BY_{i}; \bdtheta^{(t-1)})d\gamma , \\
	&& f(\gamma, L_{ic}=1|\BX_{i}, \BY_{i}; \bdtheta^{(t-1)}) =  {f(\BY_{i}|\BX_{i},\gamma;\bdtheta_y^{(t-1)})\frac{1}{\sigma_c^{(t-1)}}\phi\{(\gamma-\mu_c^{(t-1)})/\sigma_c^{(t-1)}\}\pi_c^{(t-1)}
	\over \sum_{c=1}^C \int f(\BY_{i}|\BX_{i},\gamma;\bdtheta_y^{(t-1)})f(\gamma|\bdtheta_{\gamma}^{(t-1)},L_{ic}=1)\pi_{c}^{(t-1)}d\gamma }.
\ese

Let $\{d_m\}_{m=1}^M$ and $\{w_m\}_{m=1}^M$ be Gauss-Hermite abscissas and weights, and denote $\gamma^{(c,m)}=\mu_c^{(t-1)}+\sqrt{2} \sigma_c^{(t-1)} d_m$.  
The Gauss-Hermite approximation for $Q(\bdtheta|\bdtheta^{(t-1}))$ is 
\begingroup\makeatletter\def\f@size{10}\check@mathfonts
\begin{eqnarray*}
	\wh {Q}(\boldsymbol{\theta}|{\boldsymbol{\theta}}^{(t-1)}) & = & \sum_{i=1}^{n}\frac{\sum_{c=1}^C\sum_{m=1}^M w_{m}\pi_{c}^{(t-1)}\log f(\boldsymbol{Y}_{i}|\boldsymbol{X}_{i},\gamma^{(c,m)};\bdtheta_y)f(\boldsymbol{Y}_{i}|\boldsymbol{X}_{i},\gamma^{(c,m)};\bdtheta_y^{(t-1)})}{\sum_{c=1}^C\sum_{m=1}^M w_{m}\pi_{c}^{(t-1)}f(\boldsymbol{Y}_{i}|\boldsymbol{X}_{i},\gamma^{(c,m)};\bdtheta_y^{(t-1)})}\\
	 &  & +\sum_{i=1}^{n}\frac{\sum_{c=1}^C\sum_{m=1}^M w_{m}\pi_{c}^{(t-1)}\left[-\frac{1}{2}\log 2\pi \sigma_{c}^{2}-\frac{1}{2}\frac{(\gamma^{(c,m)}-\mu_{c})^{2}}{\sigma_{c}^{2}}\right]f(\boldsymbol{Y}_{i}|\boldsymbol{X}_{i},\gamma^{(c,m)};\bdtheta_y^{(t-1)})}{\sum_{c=1}^C\sum_{m=1}^M w_{m}\pi_{c}^{(t-1)}f(\boldsymbol{Y}_{i}|\boldsymbol{X}_{i},\gamma^{(c,m)};\bdtheta_y^{(t-1)})}\\
	 &  & +\sum_{i=1}^{n}\frac{\sum_{c=1}^C\sum_{m=1}^M w_{m}\pi_{c}^{(t-1)}\log\pi_{c}f(\boldsymbol{Y}_{i}|\boldsymbol{X}_{i},\gamma^{(c,m)};\bdtheta_y^{(t-1)})}{\sum_{c=1}^C\sum_{m=1}^M w_{m}\pi_{c}^{(t-1)}f(\boldsymbol{Y}_{i}|\boldsymbol{X}_{i},\gamma^{(c,m)};\bdtheta_y^{(t-1)})} +\sum_{c=1}^Cp_n(\sigma^{2}_{c};\wh\sigma^2_{pilot}) \\
	 &=&\sum_{i=1}^{n}\sum_{c=1}^{C}  \sum_{m=1}^{M} \omega_{icm} \left\{ \log f(\BY_{i}|\BX_{i},\gamma^{(c,m)};\bdtheta_y)
  -\frac{1}{2}\log 2\pi\sigma_{c}^{2}-\frac{1}{2}\frac{(\gamma^{(c,m)}-\mu_{c})^{2}}{\sigma_{c}^{2}}
  +\log\pi_{c}  \right\}\\
  	&&-a_n \sum_{c=1}^C\{ \wh\sigma^2_{pilot}/\sigma_c^2+\log(\sigma_c^2/\wh\sigma_{pilot}^2)-1\} ,
\end{eqnarray*}
\endgroup
where
$$
\omega_{icm}=\frac{w_{m}\pi_{c}^{(t-1)}f(\boldsymbol{Y}_{i}|\boldsymbol{X}_{i},\gamma^{(c,m)};\bdtheta_y^{(t-1)})}{\sum_{c=1}^C\sum_{m=1}^M w_{m}\pi_{c}^{(t-1)}f(\boldsymbol{Y}_{i}|\boldsymbol{X}_{i},\gamma^{(c,m)};\bdtheta_y^{(t-1)})}
$$ 
as defined in (\ref{eq:posterior_weights}). Maximizing $\wh Q(\bdtheta|\bdtheta^{(t-1)})$ with respect to different components of $\bdtheta$ results in the updating scheme in Section \ref{estimation}.

%
%
%

\end{document}